\documentclass[%
  fleqn,colorlinks,linkcolor=blue,citecolor=blue,urlcolor=blue]{eptcs}
\usepackage[latin1]{inputenc}
\usepackage{amsfonts}
\usepackage{amsmath}
\usepackage{amssymb}
\usepackage{amsthm}
\usepackage{newpxtext,newpxmath}
\usepackage{z-eves}
\usepackage{multicol}
\usepackage{algorithm}
\usepackage{algpseudocode}
\usepackage{comment}
\usepackage{listings}
\usepackage{xspace}
\usepackage{fancyvrb}
\usepackage{tabularx}
\usepackage{framed}
\usepackage{enumitem}
\usepackage{soul}
\usepackage{tcolorbox}
\usepackage{tikz}
\usetikzlibrary{positioning,fit,calc,babel,decorations.text,patterns,
                shapes,arrows,automata,decorations.pathmorphing,
                shadows.blur}

\newtheorem*{teorema*}{Teorema}

\newcommand{\card}[1]{\lvert #1 \rvert}

\newcommand{\setlog}{$\{log\}$\xspace}

\newcommand{\Size}{size}

\newcommand{\Rel}{rel}

\newcommand{\Dom}{dom}
\newcommand{\Comp}{comp}
\newcommand{\set}{set}
\renewcommand{\Cup}{un}
\renewcommand{\Cap}{inters}
\newcommand{\disj}{\parallel}
\newcommand{\Diff}{diff}
\newcommand{\In}{in\xspace}
\newcommand{\Disj}{disj}
\newcommand{\Subseteq}{subset}
\renewcommand{\Subset}{ssubset}
\newcommand{\Neq}{neq\xspace}

\newcommand{\Ncup}{nun}
\newcommand{\Ncap}{ninters}
\newcommand{\Ndiff}{ndiff}
\newcommand{\Nin}{nin\xspace}
\newcommand{\Ndisj}{ndisj}
\newcommand{\Nsubseteq}{nsubset}

\newcommand{\Ndom}{ndom}
\newcommand{\Ran}{ran}

\newcommand{\Inv}{inv}

\newcommand{\Dres}{dres}
\newcommand{\Rres}{rres}
\newcommand{\Ndres}{dares}
\newcommand{\Nrres}{rares}
\newcommand{\Rimg}{rimg}
\newcommand{\Oplus}{oplus}
\newcommand{\Apply}{apply}

\newcommand{\Pfun}{pfun}

\newcommand{\List}{slist}
\newcommand{\Head}{head}
\newcommand{\Tail}{tail}
\newcommand{\Last}{last}
\newcommand{\Front}{front}
\newcommand{\Add}{add}
\newcommand{\Concat}{concat}
\newcommand{\Filter}{filter}
\newcommand{\Extract}{extract}

\newcommand{\bmachine}{\textsc{Machine}\xspace}
\newcommand{\bsets}{\textsc{Sets}\xspace}
\newcommand{\bvars}{\textsc{Variables}\xspace}
\newcommand{\binv}{\textsc{Invariant}\xspace}
\newcommand{\binit}{\textsc{Initialization}\xspace}
\newcommand{\bopers}{\textsc{Operations}\xspace}
\newcommand{\bpre}{{\footnotesize\textbf{PRE}}\xspace}
\newcommand{\bpost}{{\footnotesize\textbf{THEN}}\xspace}

\newcommand{\bendm}{\textsc{End}\xspace}

\newcommand{\bif}{{\footnotesize\textbf{IF}}~~~}
\newcommand{\bite}[3]{\bif #1 ~~~\bthen #2 ~~~\belse #3 ~~~\bend}

\newcommand{\bifp}{{\footnotesize\textbf{IF}}\xspace}
\newcommand{\bthen}{{\footnotesize\textbf{THEN}}~~~}
\newcommand{\bthenp}{{\footnotesize\textbf{THEN}}\xspace}
\newcommand{\belse}{{\footnotesize\textbf{ELSE}}~~~}
\newcommand{\bend}{{\footnotesize\textbf{END}}}
\newcommand{\belsep}{{\footnotesize\textbf{ELSE}}\xspace}

\newcommand{\SETS}[1]{\bsets~~~$#1$}
\newcommand{\VARS}[1]{\bvars~~~$#1$}
\newcommand{\INV}[1]{\binv~~~$#1$}
\newcommand{\INIT}[1]{\binit~~~$#1$}
\newcommand{\OPERS}{\bopers}

\newcommand{\OperO}[3]{~~~~~~$#3 \leftarrow \textbf{#1}(#2) \defs$}
\newcommand{\Pre}[1]{~~~~~~~~~\bpre~~~$#1$}
\newcommand{\Post}[1]{~~~~~~~~~\bpost~~~$#1$}
\newcommand{\EndOp}{~~~~~~~~~\bend}
\newenvironment{machine}[1]%
  {\begin{tabular}{l}
   \mbox{} \\ 
   \bmachine~~~\textit{#1} \\
  }
  {\bendm \\ \mbox{} \end{tabular}}

\date{}

\begin{document}

\DefineShortVerb[commandchars=\\\$\$]{\@}

\thispagestyle{empty}
\begin{center}
{\bf
Class notes \\\vfill
{\Huge From B Specifications to \setlog Forgrams} \vfill
{\large Maximiliano Cristiá} \\\vfill
Computational Science 3 \\[5mm]
Bachelor in Computer Science \\[5mm]
Faculty of Science, Technology and Medicine  \\[5mm]
University of Luxembourg} \\\vfill
{\footnotesize
\copyright\ Maximiliano Cristiá -- 2023 -- All rights reserved \\
These class notes have been written during a visit to University of Luxembourg from February until June, 2023.
}
\end{center}

\pagebreak

\thispagestyle{empty}
\tableofcontents
\vfill

\pagebreak

\section{What is \setlog?}
\setlog (`setlog') is a constraint logic programming language. Besides it's a \emph{satisfiability solver} and as such it can be used as an automated theorem prover. One of \setlog's distinctive features is that sets are first-class entities of the language.

\setlog was first developed by Gianfranco Rossi and his PhD students in Italy during the mid '90. Since 2012 Gianfranco Rossi and Maximiliano Cristiá work together in extending \setlog in different directions. 

As shown below, \setlog is at the intersection of several Computer Science areas. \setlog can be used as a \href{https://en.wikipedia.org/wiki/Formal_verification}{\emph{formal verification}} tool because it performs \href{https://en.wikipedia.org/wiki/Automated_theorem_proving}{\emph{automated proofs}} over a very expressive theory. It's also a \href{https://en.wikipedia.org/wiki/Declarative_programming}{\emph{declarative programming language}} meaning that programmers have to expresses the logic of a computation without describing its control flow. In particular, \setlog implements declarative programming as an instance of a \href{https://en.wikipedia.org/wiki/Constraint_logic_programming}{\emph{constraint logic programming}} (CLP) system implemented in \href{https://en.wikipedia.org/wiki/Prolog}{Prolog}. The code written in \setlog is quite similar (in its essence, not in its form) to formal specifications written in languages based on set theory and set relation algebra such as B, \href{https://en.wikipedia.org/wiki/Z_notation}{Z} and \href{https://en.wikipedia.org/wiki/Alloy_(specification_language)}{Alloy}.

\centerline{%
\includegraphics[clip,trim=5cm 15.5cm 5cm 4cm]{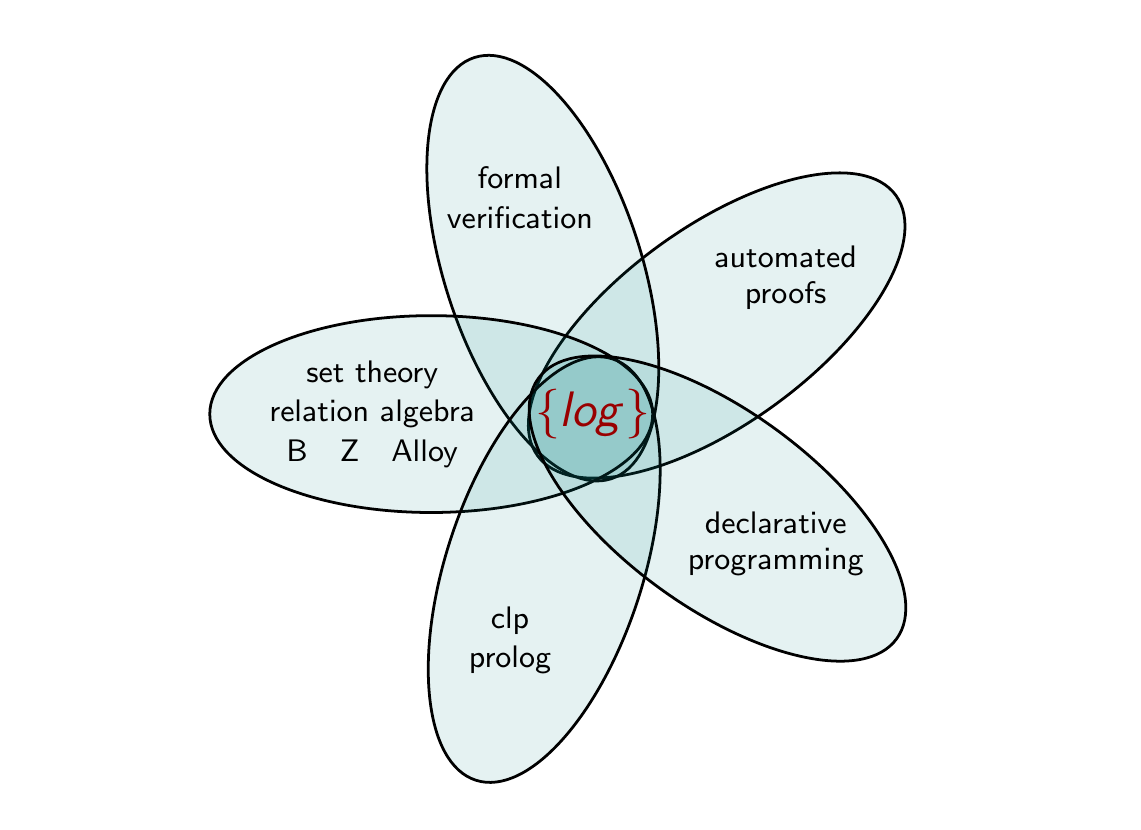}
}

\subsection{Installation}
\setlog is a Prolog program. Then, you first need to install a Prolog interpreter. So far \setlog runs only on SWI-Prolog (\url{http://www.swi-prolog.org}). After installing SWI-Prolog you must download  \setlog, all the library files and its user's manual from here:

\centerline{\url{https://www.clpset.unipr.it/setlog.Home.html}}

You should also read \setlog user's manual:

\centerline{\url{https://www.clpset.unipr.it/SETLOG/setlog-man.pdf}}

\subsection{Using \setlog}
As we have said, \setlog is a \emph{satisfiability solver}. This means that \setlog is a program that determines whether or not a given formula is satisfiable. Once yo access \setlog it presents a prompt:
\begin{Verbatim}
{log}=>
\end{Verbatim}
You can now ask \setlog to solve formulas. For example:
\begin{Verbatim}
{log}=> un({a,2},B,{X,2,c}).
\end{Verbatim}
The atomic predicate \verb+un({a,2},B,{X,2,c})+ means $\{a,2\} \cup B = \{X,2,c\}$, where $X$ and $B$ are variables and $a$ and $c$ are constants. In \setlog variables begin with a uppercase letter, and constants begin with lowercase letters. Note that the formula ends with a dot. Hence, when we type in that formula \setlog will try to find values for \verb+B+ and \verb+X+ that satisfy the formula---this is why we say that \setlog is a satisfiability solver. So, \setlog asks itself, are there values for \verb+B+ and \verb+X+ that make the formula true? \setlog answers the following:
\begin{Verbatim}
B = {c},  
X = a

Another solution?  (y/n)
\end{Verbatim}
As you can see, \setlog produces a solution and asks whether or not we want to see other solutions. In this case there are three more solutions:
\begin{Verbatim}
B = {2,c},  
X = a

Another solution?  (y/n)
B = {a,c},  
X = a

Another solution?  (y/n)
B = {a,2,c},  
X = a

Another solution?  (y/n)
no

{log}=>
\end{Verbatim}
When there are no more solutions or when we don't type in `\verb+y+', \setlog says `\verb+no+' and prints the prompt again.

Let's try another example.
\begin{Verbatim}
{log}=> un({a,2},B,{X,2,c}) & c nin B.
\end{Verbatim}
The atomic predicate \verb+c nin B+ means $c \notin B$ and `\verb+&+' means conjunction ($\land$). In this case \setlog answers \verb+no+. Why is that? Because there are no values for \verb+B+ and \verb+X+ that make the formula true. Clearly, as \verb+c+ doesn't belong to \verb+{a,2}+ but at the same time it belongs to the union between that set and \verb+B+ the only chance to satisfy the formula is when \verb+c+ belongs to \verb+B+. But we rule this possibility out by conjoining \verb+c nin B+. Then, \setlog is saying ``your formula is unsatisfiable''.

Summarizing, if we see anything different from `\verb+no+' we know the formula is satisfiable; otherwise, it's unsatisfiable.

\section{\label{traduccion}An example of a B specification translated into \setlog}
These class notes are focused in showing how B specifications can be translated into \setlog and, later, on how \setlog can be used to run simulations and automated proofs.

Many B specifications can be easily translated into \setlog. This means that \setlog can serve as a programming language in which a prototype of a B specification can be immediately implemented. 

We have already learned to write some B specifications. Here, we will show how these B specifications can be translated into \setlog. To that end we will use a running example. Later on we will explain with some detail how B elements not appearing in the example can be translated into \setlog; we will see that some B elements can be translated in more than one way.

\subsection{The running example}
The specification to be used as running example is known as the \emph{birthday book}. It's a system which records people's birthdays, and is
able to issue a reminder when the day comes round. The problem is borrowed from \cite{Spivey00}.

\subsection{The B specification}
The B machine containing the specification of the birthday book system will be called $BirthdayBook$. In our account of the system, we need to deal with people's names and
with dates. We also need a type for the messages outputted by some of the operations. Then, we introduce the following types.

\begin{machine}{BirthdayBook}
\SETS{NAME; DATE; MSG = \{ok, nameExists\}} \\
\dots\dots\dots\\
\end{machine}

Now, we define two state variables for our machine:

\begin{machine}{BirthdayBook}
\SETS{NAME; DATE; MSG = \{ok, nameExists\}} \\
\VARS{known, birthday} \\
\dots\dots\dots\\
\end{machine} \\
where $known$ is the set of names with birthdays recorded; and $birthday$ is a function which, when applied to certain names, gives the birthdays associated with them.

The invariant of our machine is the following.

\begin{machine}{BirthdayBook}
\SETS{NAME; DATE; MSG = \{ok, nameExists\}} \\
\VARS{known, birthday} \\
\INV{known \in \power NAME \land birthday \in NAME \pfun DATE \land known = \dom(birthday)} \\
\dots\dots\dots\\
\end{machine} \\
As can be seen, the value of $known$ can be derived from the value of $birthday$. This makes $known$ a
\emph{derived} component.  It would be possible to specify the system without mentioning $known$ at all.
However, giving names to important concepts helps to make specifications more
readable. The specification doesn't commit the programmer to represent $known$ explicitly in an implementation. Besides the types for the variables are in accordance with the intended use described above.

The initial state of the birthday book is the following.

\begin{machine}{BirthdayBook}
\SETS{NAME; DATE; MSG = \{ok, nameExists\}} \\
\VARS{known, birthday} \\
\INV{known \in \power NAME \land birthday \in NAME \pfun DATE \land known = \dom(birthday)} \\
\INIT{known, birthday := \{\}, \{\}} \\
\dots\dots\dots\\
\end{machine}

The first operation we specify is how to add a birthday to the birthday book. As we did with the savings account specification we model the normal and abnormal behaviors outputting convenient messages in each case.
\[
msg \leftarrow \textbf{addBirthday} (name, date) \defs \\
\t1\bpre~~~ name \in NAME \land date \in DATE \\
\t1\bthen \\
  \t2 \bif name \notin known \\
  \t2\bthen known, birthday, msg := known \cup \{name\}, birthday \cup \{name \mapsto date\}, ok \\
  \t2\belse msg := nameExists \\
  \t2\bend \\
\t1 \bend
\]
Note how both state variables are updated accordingly.

The second operation to be specified is the one that shows the birthday of a given person.
\[
date \leftarrow \textbf{findBirthday} (name) \defs \\
\t1\bpre~~~ name \in NAME \land name \in known \\
\t1\bthen date := birthday(name) \\
\t1 \bend
\]

Finally we have an operation listing all the persons whose birthday is a given date.
\[
cards \leftarrow \textbf{remind} (today) \defs \\
\t1\bpre~~~ today \in DATE \\
\t1\bthen cards := \dom(birthday \rres \{today\}) \\
\t1 \bend
\]

The complete B specification of the birthday book can be seen in Figure \ref{fig:birthdaybook}.

\begin{figure}[t]
\begin{machine}{BirthdayBook}
\SETS{NAME; DATE; MSG = \{ok, nameExists\}} \\
\VARS{known, birthday} \\
\INV{known \in \power NAME \land birthday \in NAME \pfun DATE \land known = \dom(birthday)} \\
\INIT{known, birthday := \{\}, \{\}} \\
\OPERS \\
\OperO{addBirthday}{name,date}{msg} \\
\Pre{name \in NAME \land date \in DATE} \\
\Post{} \\
\t2 \bif $name \notin known$ \\
\t2 \bthen $known, birthday, msg := known \cup \{name\}, birthday \cup \{name \mapsto date\}, ok$ \\
\t2 \belse $msg := nameExists$ \\
\t2 \bend \\
\EndOp; \\
\OperO{findBirthday}{name}{date} \\
\Pre{name \in NAME \land name \in known} \\
\Post{date := birthday(name)} \\
\EndOp; \\
\OperO{remind}{today}{cards} \\
\Pre{today \in DATE} \\
\Post{cards := \dom(birthday \rres \{today\})} \\
\EndOp \\
\end{machine}
\caption{\label{fig:birthdaybook} B specification of the birthday book}
\end{figure}

\subsection{\label{codigosetlog}The \setlog forgram}
The \setlog forgram resulting from the translation of the B specification must be saved in a file with extension \Verb+.pl+ or \Verb+.slog+. It is convenient to put this file in the same folder where \setlog was installed.

A B machine is translated as a collection of \setlog \emph{clauses} and \emph{declarations} written in a single file. A \setlog clause is a sort of subroutine or subprogram or procedure of a regular programming language. Each clause can receive zero or more arguments. In \setlog variables must always begin with an uppercase letter or the underscore character (\Verb+_+), although this is usually saved for special cases. Any identifier beginning with a lowercase letter is a constant. Then, for instance, the state variables of the birthday book will be \Verb+Known+ and \Verb+Birthday+, instead of \Verb+known+ and \Verb+birthday+ because in this case they would be constants. We'll see how variables are typed in Section \ref{types}. For now we'll not pay much attention to types.

\subsubsection{Translating the \bsets section} 
In general, the \bsets sections is not translated into \setlog. The sets declared in this section can be freely introduced in \setlog. We'll see more on this in Section \ref{types}.

\subsubsection{Translating the \bvars section}
The \bvars section is translated as a \setlog declaration as follows:
\begin{Verbatim}
variables([Known, Birthday]).
\end{Verbatim}

Note that declarations end with a dot (`\verb+.+').

\subsubsection{\label{inv}Translating the \binv section}
Before translating the invariant we normalize it:
\[
\binv~~~ known \in \power NAME \land birthday \in NAME \rel DATE \\
\t3 \land pfun(birthday) \land known = \dom(birthday)
\]
The first part of the invariant ($known \in \power NAME \land birthday \in NAME \rel DATE$) is translated as type declarations, whereas the second part is translated as a clause declared as invariant. Type declarations will be introduced in Section \ref{types}. The \setlog code is the following:
\begin{Verbatim}
invariant(birthdayBookInv).
birthdayBookInv(Known,Birthday) :- dom(Birthday,Known) & pfun(Birthday).
\end{Verbatim}
Then, the first line declares the clause named \verb+birthdayBookInv+ to be an invariant. The second line is a clause. 

Clauses are of the form:
\begin{Verbatim}
head(params) :- body.
\end{Verbatim}
where \verb+body+ is a \setlog formula. In this case the formula is simply \verb+dom(Birthday,Known) & pfun(Birthday)+ which is equivalent to $known = \dom(birthday) \land pfun(birthday)$. 

Alternatively, you can split the invariant in smaller pieces. Actually, each conjunct in the \binv section may become an invariant. This strategy is a good option when the specification is large and complex  because later it will be easier for \setlog to discharge invariance lemmas. In this case the \setlog code look like this:
\begin{Verbatim}
invariant(birthdayBookInv).
birthdayBookInv(Known,Birthday) :- dom(Birthday,Known).

invariant(pfunInv).
pfunInv(Birthday) :- pfun(Birthday).
\end{Verbatim}

Note that declarations and clauses end with a dot (`\verb+.+').

\subsubsection{Translating the \binit section}
The \binit section is translated as a declaration and a clause as follows:
\begin{Verbatim}
initial(birthdayBookInit).
birthdayBookInit(Known,Birthday) :- Known = {} & Birthday = {}.
\end{Verbatim}
That is, we first declare that the clause \verb+birthdayBookInit+ corresponds to the initial state of the system and then the clause is defined. Here there's an important difference w.r.t. the B specification because the body of the clause is a formula and not a multiple assignment. Indeed, \verb+Known = {}+ and \verb+Birthday = {}+ are predicates. We could have written them also as \verb+{} = Known+ and \verb+{} = Birthday+ because the symbol `\verb+=+' is simply logical equality. In turn `\verb+&+' means conjunction ($\land$). Hence, we could have written \verb+birthdayBookInit+ as follows:
\begin{Verbatim}
birthdayBookInit(Known,Birthday) :- {} = Birthday & {} = Known.
\end{Verbatim}

In any case, the \setlog implementation of the \binit section follows the semantics of the B specification.

\subsubsection{Translating operations}
A B operation is translated as a clause and a declaration indicating that the clause is an operation. When a B operation is translated, the corresponding clause receives as arguments all the state variables, all the input parameters and all the output parameters. Besides, for each state variable $v$ the clause will also receive $v\_$, which represents the value of $v$ in the next state. That is, in \setlog we have to represent the next state explicitly with a second set of variables. 
Hence, the head of the \setlog clause corresponding to the B operation named \textbf{addBirthday} is the following:
\begin{Verbatim}
addBirthday(Known,Birthday,Name,Date,Known_,Birthday_,Msg)
\end{Verbatim}
where \Verb+Name+ and \Verb+Date+ correspond to input parameters $name$ and $date$ declared in \textbf{addBirthday}; \Verb+Known+ and \Verb+Birthday+ represent the before state while \Verb+Known_+ and \Verb+Birthday_+ represent the after state; and \verb+Msg+ corresponds to the output parameter.

Now we give the complete specification of the clause preceded by its declarion:
\begin{Verbatim}
operation(addBirthday).
addBirthday(Known,Birthday,Name,Date,Known_,Birthday_,Msg) :-
  (Name nin Known &
   un(Known,{Name},Known_) &
   un(Birthday,{[Name,Date]},Birthday_) &
   Msg = ok
  or
   Name in Known &
   Known_ = Known &
   Birthday_ = Birthday &
   Msg = nameExists
  ).
\end{Verbatim}
That is, the first line declares that \verb+addBirthday+ is an operation. Then, the \bifp-\bthenp-\belsep statement in \textbf{addBirthday} is translated as a logical disjunction (`\verb+or+'). The condition of the conditional statement, $name \notin known$, is translated as \verb+Name nin Known+. The word `\verb+nin+' in \setlog means $\notin$. If the condition is true the \bthenp branch specifies the multi assignment:
\[
known, birthday, msg := known \cup \{name\}, birthday \cup \{name \mapsto date\}, ok
\]
This multi assignment is translated as a conjunction of \setlog constraints:
\begin{Verbatim}
un(Known,{Name},Known_) & un(Birthday,{[Name,Date]},Birthday_) & Msg = ok
\end{Verbatim}
The meaning of these constraints is as follows:
\begin{itemize}
\item \Verb+un(Known,{Name_i},Known_)+ means $Known\_ = Known \cup \{Name\}$.

That is, in \setlog \verb+un(A,B,C)+ is equivalent to $C = A \cup B$.
\item Similarly, \verb+un(Birthday,{[Name,Date]},Birthday_)+ is $Birthday\_ = Birthday \cup \{Name \mapsto Date\}$.

That is, in \setlog the ordered pair $x \mapsto y$ is written as \verb+[x,y]+.
\end{itemize}

When the condition of the \bifp-\bthenp-\belsep statement is false, we have the assignment $msg := nameExists$. This means that the state of the machine doesn't change and that the machine outputs $nameExists$. In \setlog we first need to write the negation of the condition, that is \verb+Name in Known+ or \verb+neg(Name nin Known)+. Then, we must say that the machine doesn't change the state and that $nameExists$ is outputted. We do this with the conjunction:
\begin{Verbatim}
Known_ = Known & Birthday_ = Birthday & Msg = nameExists
\end{Verbatim}
As \verb+Known_+ and \verb+Birthday_+ represent the next state, the equalities \verb+Known_ = Known+ and \verb+Birthday_ = Birthday+ mean that the state doesn't change.

Finally, observe that the \bpre section hasn't been translated. In this case the \bpre section contains only type declarations ($name \in NAME \land date \in DATE$). The translation of type declarations will be seen in Section \ref{types}.

Now we give the translation of \textbf{findBirthday}.
\begin{Verbatim}
operation(findBirthday).
findBirthday(Known,Birthday,Name,Date,Known,Birthday) :-
  Name in Known & applyTo(Birthday,Name,Date).
\end{Verbatim}
where \Verb+applyTo+ is a predicate implementing function application. That is, \Verb+applyTo(F,X,Y)+ is true if and only if $F(X) = Y$ holds. Note that \Verb+applyTo(F,X,Y)+ makes sense only if \verb+X+ is in the domain of \Verb+F+, which in turn is a function \emph{at least on} \verb+X+. As with \textbf{addBirthday} the type declaration $name \in NAME$ isn't included in the body of the clause. Besides, note how we say that the operation doesn't change the state. Instead of including \verb+Known_ = Known & Birthday_ = Birthday+ in the body of the clause we don't include \verb+Known_+ and \verb+Birthday_+ in the head but two copies of the before-state variables. This is interpreted by \setlog as the operation not changing the state. We couldn't do this in \textbf{addBirthday} because there's one branch of that operation that changes the state.

Finally, the translation of \textbf{remind} is the following:
\begin{Verbatim}
operation(remind).
remind(Known,Birthday,Today,Cards,Known,Birthday) :-
  rres(Birthday,{Today},M) & dom(M,Cards).
\end{Verbatim}
This is an interesting example because it shows how set and relational expressions must be translated. Given that in \setlog set and relational operators are implemented as predicates, it's impossible to write set and relational expressions. Instead, we have to introduce new variables (such as \Verb+M+) to ``chain'' the predicates. Predicate \Verb+rres(R,A,S)+ stands for $S = R \rres A$. Then, the body of the clause corresponds to the following B predicate: $m = birthday \rres \{today\} \land cards = dom(m)$. As \textbf{remind} doesn't change the state we repeat the state variables in the head of the clause.

\section{\label{types}Types in \setlog}
So far we haven't given the types of the variables. \setlog provides a typechecker that can be activated and deactivated by the user.  \setlog's type system is described in detail in chapter 9 of \setlog user's manual. Here we will give a broad description of how to use types in \setlog.

\setlog's type system allows users to define type synonyms to simplify the type declaration of clauses and variables. For example, we can define the following type synonyms for the birthday book:
\begin{Verbatim}
def_type(bb,rel(name,date)).
def_type(kn,set(name)).
def_type(msg,enum([ok,nameExists])).
\end{Verbatim}
where \Verb+bb+ is a type identifier o synonym of the type \Verb+rel(name,date)+. In \Verb+rel(name,date)+, \Verb+name+ and \Verb+date+ correspond to the basic types $NAME$ and $DATE$ of the B specification. B basic types can be introduced in \setlog without any previous declaration. In \setlog basic types must begin with a lowercase letter (i.e. they are constants). In turn, \Verb+rel(name,date)+ corresponds to the type of all binary relations between \Verb+name+ and \Verb+date+. That is, \Verb+rel(name,date)+ corresponds to $NAME \rel DATE$ in B. \verb+set(name)+ corresponds to $\power NAME$ in B and \verb+enum([ok,nameExists])+ corresponds to the set $\{ok, nameExists\}$ which we named $MSG$ in the B specification.

These type synonyms allow us to declare the type of the  \Verb+addBirthday+ operation:
\begin{Verbatim}
dec_p_type(addBirthday(kn,bb,name,date,kn,bb,msg)).
\end{Verbatim}
The type declaration must come before the clause definition:
\begin{Verbatim}
operation(addBirthday).
dec_p_type(addBirthday(kn,bb,name,date,kn,bb,msg)).
addBirthday(Known,Birthday,Name,Date,Known_,Birthday_,Msg) :-
  (Name nin Known &
  ...
\end{Verbatim}

The \Verb+dec_p_type+ declaration has only one argument of the following form:
\begin{Verbatim}
clause_name(parameters)
\end{Verbatim}
In turn, \Verb+parameters+ is a list whose elements corresponds one-to-one to the clause arguments. Then, the type of \Verb+Known+ is \Verb+kn+, the type of \Verb+Birthday+ is \Verb+bb+, etc. 

The following is the typed version of the \Verb+remid+ operation.
\begin{Verbatim}
operation(remind).
dec_p_type(remind(kn,bb,date,kn,kn,bb)).
remind(Known,Birthday,Today,Cards,Known,Birthday) :-
  rres(Birthday,{Today},M) & dom(M,Cards) & dec(M,bb).
\end{Verbatim}
This clause is interesting because it shows how variables local to the clause are typed by means of the \Verb+dec(V,t)+ predicate. Indeed, \Verb+dec(V,t)+ is interpreted as ``variable \Verb+V+ is of tye \Verb+t+''.

The \setlog forgram including type declarations of the complete translation of the birthday book can be found in Appendix \ref{ap:bb}. As can be seen in that appendix, all the clauses, including \verb+invariant+ and \verb+initial+, are typed.

Recall that partial functions \emph{aren't} a type in B. The same happens in \setlog; in fact it is impossible to define the type of all partial functions. The natural numbers are another example of a set that isn't a type. This means that if in B we have $f \in X \pfun Y$ in \setlog we declare \Verb+F+ to be of type \Verb+rel(x,y)+ and then we should prove that \Verb+F+ is a function as an invariant. Likewise, if in B we declare $x \in \nat$, in \setlog we must declare \Verb+X+ to be of type \Verb+int+ and then prove that \Verb+0 =< X+ is an invariant. In general, when a B specification is translated into \setlog it is convenient to first normalize the B specification and then start the translation into \setlog. In this case B types are translated straightforwardly and the predicates introduced due to the normalization process become constraints at the \setlog level (i.e. \Verb+0 =< X+) or they are proved to be invariants. For instance, $x \in \nat$ is a non-normalized declaration because $\nat$ isn't a type (it's a set). The normalized declaration is $x \in \num$ plus $x \geq 0$ conjoined in the \binv section or in the \bpre section of an operation. In this case, in \setlog the type of $x$ is \Verb+int+ and we should prove that $x$ is always greater than or equal to zero (i.e., that $0 \leq x$ is an invariant), or simply assert that as a precondition.

\section{Translating B specifications into \setlog}
In this section we show how the most used elements appearing in B specification are translated into \setlog.

\subsection{\label{aritmetica}Translating arithmetic expressions}
Almost all Z arithmetic expressions are translated directly into \setlog, with some exceptions. The relational symbols $\leq$, $\geq$ and $\neq$ are translated as \Verb+=<+, \Verb+>=+ and \Verb+neq+, respectively. The arithmetic operators are the usual ones: \Verb.+., \Verb.-., \Verb.*., \Verb.div. y \Verb.mod..  

An equality of the form $x' = x + 1$ is translated as \Verb!X_ is X + 1! (that is, in arithmetic equalities you mustn't use `\Verb+=+' but `\Verb+is+'). Furthermore, if in B we have $A = \{x, y - 4\}$ ($A$, $x$ and $y$ variables) it has to be encoded as: \Verb+A = {X,Z} & Z is Y - 4+, where \Verb+Z+ is a variable not used in the clause. The problem is that \setlog doesn't evaluate arithmetic expressions unless the programmer forces it by using the \Verb+is+ operator. This means that if in \setlog we run \Verb+{X,Y - 4} = {Y - 3 - 1,X}+, the answer will be \Verb+no+ because \setlog will try to find out whether or not \Verb+Y - 4 = Y - 3 - 1+ without evaluating the expressions (that is, it will consider them, basically, as character strings where \Verb+Y+ is an integer variable and thus it is impossible for the equality to hold regardless of the value of \Verb+Y+). On the contrary, if we run \Verb+{X,A} = {B,X} & A is Y - 4 & B is Y - 3 - 1+ \setlog will return several solutions (with some repetitions), meaning that the sets are equal in several ways.

The same applies to the \Verb+neq+ predicate: for \setlog \Verb+Y - 4 neq Y - 3 - 1+ is true. As a consequence we must write: \Verb+H is Y - 4 & U is Y - 3 - 1 & H neq U+. However, this is not necessary with the order predicates: \Verb.X + 1 > X. is satisfiable but \Verb+X - 1 > X+ isn't.

\subsection{Translating ordered pairs}
Ordered pairs are encoded as Prolog lists of two elements. For instance, if $x$ is a variable $(x,3)$ or $x \mapsto 3$ is translated as \Verb+[X,3]+.

If in B we have $p \in X \cross Y$ then the \setlog type declaration for $p$ is \verb+dec(P,[x,y])+, where \verb+x+ corresponds to the encoding of type $X$ in \setlog; similarly for \verb+y+.

\subsection{Translating sets}

\subsubsection{Extensional sets --- Introduction to set unification}
In \setlog the empty set is written as in B, \verb+{}+. The set $\{1,2,3\}$ is simply translated as \Verb+{1,2,3}+. If one of the elements of the set is a variable or an element of an enumerated type, take care of the differences concerning variables and constants in B and \setlog. For example, if in B $x$ is a variable, then the set $\{2,x,6\}$ is translated as \Verb+{2,X,6}+; and if in B $Run$ is an element of a set declared in the \bsets section, then the set $\{2,Run,6\}$ is translated as \Verb+{2,run,6}+.

However, \setlog provides a form of extensional sets that, in a sense, is more powerful than the one offered in B. The term \Verb+{.../...}+ is called \emph{extensional set constructor}. In \Verb+{E/C}+ the second argument (i.e. \Verb+C+) must be a set. \Verb+{E/C}+ means $\{E\} \cup C$. Then, there are solutions where $E \in C$. To avoid such solutions (in case they're incorrect or unwanted) the predicate $E \notin C$ must be explicitly added to the formula. In order to make the language more simple, \setlog accepts and prints terms such as  \Verb+{1,2 / X}+ instead of \Verb+{1 / {2 / X}}+.

The extensional set constructor is useful and in general it's more efficient than other encodings. For example, the B assignment (assume $d$ is a variable):
\[
A := A \setminus \{d\}
\]
can be translated by means of the \setlog predicate \Verb+diff+, whose semantics is equivalent to $\setminus$ (see Table \ref{t:setoper}):
\begin{Verbatim}
diff(A,{D},A_)
\end{Verbatim}
Bu it also can be translated by means of an extensional set:
\begin{Verbatim}
A = {D / A_} & D nin A_ or D nin A & A_ = A
\end{Verbatim}
which in general is more efficient.

That is, the predicate \Verb+A = {D / A_}+ \emph{unifies} \Verb+A+ with \Verb+{D / A_}+ in such a way that it finds values for the variables to make the equality true. If such values don't exist the unification fails and \setlog tries the second disjunct. 

Why we conjoined \Verb+D nin A_+? Simply because, for instance, \Verb+A = {1,2}+, \Verb+D = 1+ and \Verb+A_ = {1,2}+ is a solution of the equation but it isn't a solution of $A := A \setminus \{d\}$. Precisely, when \Verb+D nin A_+ is conjoined all the solutions where \Verb+D+ belongs to \Verb+A_+ are eliminated.

\setlog solves equalities of the form \Verb+B = C+, where \Verb+B+ and \Verb+C+ are terms denoting sets, by using \emph{set unification}. Se unification is at the base of the deductive power of \setlog making it an important extension of Prolog's unification algorithm. Set unification is inherently computationally hard because finding out whether or not two sets are equal implies, in the worst case, computing all the permutations of their elements. On top of that, it is the fact that \setlog can deal with \emph{partially specified} sets, that is sets where some of their elements or part of the set are variables. For these reasons, in general, \setlog will show efficiency problems when dealing with certain formulas but, at the same time, we aren't aware of other tools capable to solve some of the problems \setlog can.

\subsubsection{Cartesian products}
In \setlog Cartesian products are written \Verb+cp(A,B)+ where \Verb+A+ and \Verb+B+ can be variables, extensional sets and Cartesian products.

\subsubsection{Integer intervals}
A B integer interval such as $m \upto n$ is translated as \Verb+int(m,n)+. \verb+m+ and \verb+n+ can be integer constants or variables. If we need to write something like $m+1 \upto 2*n+3$ we do as follows: \verb.int(K,J) & K is M + 1 & J is 2*N + 3., where \verb+K+ and \verb+J+ must be new variables.

\subsection{Translating set and relational operators}
Set, relational, functional and sequence operators are translated as shown in Tables \ref{t:setoper}, \ref{t:reloper} and \ref{t:listoper}.

In order to be able to work with the sequence operators shown in Table \ref{t:listoper} load the corresponding library file (e.g. \Verb+consult('setlogliblist.slog')+) into the \setlog environment. 

\begin{table}
\begin{tabularx}{\textwidth}{Xll}
\hline\hline
\multicolumn{1}{c}{\textsc{Operator}} &
\multicolumn{1}{c}{\setlog} &
\multicolumn{1}{c}{\textsc{Meaning}} \\\hline
set           & \Verb+\set(A)+        & $A$ is a set \\
equality             & \Verb+A = B+          & $A = B$ \\
set membership          & \Verb+x \In A+        & $x \in A$ \\
union                & \Verb+\Cup(A,B,C)+    & $C = A \cup B$ \\
intersection         & \Verb+\Cap(A,B,C)+    & $C = A \cap B$ \\
difference         & \Verb+\Diff(A,B,C)+   & $C = A \setminus B$ \\
subset          & \Verb+\Subseteq(A,B)+ & $A \subseteq B$ \\
strict subset & \Verb+\Subset(A,B)+   & $A \subset B$ \\
disjointness  & \Verb+\Disj(A,B) +    & $A \disj B$ \\
cardinality         & \Verb+\Size(A,n)+     & $\card{A} = n$ \\
\hline
\multicolumn{3}{c}{\textsc{Negations}} \\\hline
%
equality      & \Verb+A \Neq B+        & $A \neq B$ \\
set membership    & \Verb+x \Nin A+        & $x \notin A$ \\
union         & \Verb+\Ncup(A,B,C)+    & $C \neq A \cup B$ \\
intersection  & \Verb+\Ncap(A,B,C)+    & $C \neq A \cap B$ \\
difference    & \Verb+\Ndiff(A,B,C)+   & $C \neq A \setminus B$ \\
subset       & \Verb+\Nsubseteq(A,B)+ & $A \not\subseteq B$ \\
disjointness & \Verb+\Ndisj(A,B)+    & $A \not\disj B$ \\
\hline\hline
\end{tabularx}
\caption{\label{t:setoper} Set operators available in \setlog}
\end{table}

\begin{table}
\begin{tabularx}{\textwidth}{Xll}
\hline\hline
\multicolumn{1}{c}{\textsc{Operator}} &
\multicolumn{1}{c}{\setlog} &
\multicolumn{1}{c}{\textsc{Meaning}} \\\hline
binary relation        & \Verb+\Rel(R)+       & $R$ is a binary relation \\
partial function         & \Verb+\Pfun(R)+ & $R$ is a partial function \\
function application   & \Verb+\Apply(f,x,y)+ & $f(x) = y$ \\
domain                 & \Verb+\Dom(R,A)+     & $\dom R = A$ \\
range                   & \Verb+\Ran(R,A)+     & $\ran R = A$ \\
composition             & \Verb+\Comp(R,S,T)+  & $T = R \circ S$ \\
inverse                 & \Verb+\Inv(R,S)+     & $S = R^{-1}$ \\
domain restriction  & \Verb+\Dres(A,R,S)+  & $S = A \dres R$ \\
domain anti-restriction & \Verb+\Ndres(A,R,S)+ & $S = A \ndres R$ \\
range restriction  & \Verb+\Rres(R,A,S)+  & $S = R \rres A$ \\
range anti-restriction  & \Verb+\Nrres(R,A,S)+ & $S = R \nrres A$ \\
update          & \Verb+\Oplus(R,S,T)+ & $T = R \oplus S$ \\
relational image      & \Verb+\Rimg(R,A,B)+  & $B = R[A]$ \\\hline
\multicolumn{3}{c}{\textsc{Negations}} \\\hline
\multicolumn{3}{l}{\parbox{.95\textwidth}{\vspace{2pt}
All negations are written by prefixing a letter \Verb+n+ to the corresponding operator. For example, the negation of \Verb+\Dom(R,A)+ is \Verb+\Ndom(R,A)+, that of \Verb+\Ndres(A,R,S)+ is \Verb+n\Ndres(A,R,S)+, etc.}}
  \\[7pt]\hline\hline
\end{tabularx}
\caption{\label{t:reloper} Relational operators available in \setlog}
\end{table}

\begin{table}
\begin{tabularx}{\textwidth}{Xll}
\hline\hline
\multicolumn{1}{c}{\textsc{Operator}} &
\multicolumn{1}{c}{\setlog} &
\multicolumn{1}{c}{\textsc{Meaning}} \\\hline
sequence         & \Verb+\List(s)+        & $s$ is a sequence \\
extensional sequence & \verb+{[1,a],[2,b],...,[n,z]}+ & $\langle a,b,\dots,z\rangle$ \\
head       & \Verb+\Head(s,e)+      & $e = head~s$ \\
tail           & \Verb+\Tail(s,t)+      & $t = tail~s$ \\
last         & \Verb+\Last(s,e)+      & $e = last~s$ \\
front         & \Verb+\Front(s,t)+     & $t = front~s$ \\
add (cons) & \Verb+\Add(s,e,t)+     & $t = s \cat \langle e \rangle$ \\
concatenation  & \Verb+\Concat(s,t,u)+  & $u = s \cat t$ \\
filter       & \Verb+\Filter(A,s,t)+  & $t = A \filter s$ \\
extraction     & \Verb+\Extract(s,A,t)+ & $t = s \extract A$ \\
\hline\hline
\end{tabularx}
\caption{\label{t:listoper} Sequence operators available in \setlog}
\end{table}

The cardinality operator accepts as second argument only a constant or a variable. Hence, if we run \Verb!size(A,X + 1)! \setlog answers \Verb+no+; instead if we run \Verb!size(A,Y) & Y is X + 1! (\Verb+Y+ must be a variable not used in the clause) the answer is \Verb+true+ because the formula is satisfiable. \setlog will answer \Verb+no+ if we execute \Verb!size(A,Y) & Y = X + 1!.

\subsection{Translating function application}
One interesting application of set unification is the application of a function to its argument. Given that partial functions are frequently used in B it's necessary to add predicates of the form $x \in \dom f$, before attempting to apply $f$ to $x$. The translation of these formulas into \setlog can be done by using the predicate \Verb+applyTo+ or by using a set membership predicate which leads to set unification. For example the B formula:
\begin{zed}
x \in \dom f \land f(x) = y
\end{zed}
can be translated in a direct fashion:
\begin{Verbatim}
dom(F,D) & X in D & applyTo(F,X,Y)
\end{Verbatim}
or just using \verb+applyTo+:
\begin{Verbatim}
applyTo(F,X,Y)
\end{Verbatim}
or using set unification (if we \emph{assume} that \verb+F+ is a function):
\begin{Verbatim}
F = {[X,Y] / G} & [X,Y] nin G
\end{Verbatim}

The \setlog definition of \verb+applyTo+ is the following:
\begin{Verbatim}
applyTo(F,X,Y) :- F = {[X,Y] / G} & [X,Y] nin G & comp({[X,X]},G,{}).
\end{Verbatim}
If we know that $x \in \dom f$ the there exist \Verb+Y+ and \Verb+G+ such that \Verb+F = {[X,Y] /+ \Verb+G} & [X,Y] nin G+. Besides, if we are saying that we can apply $f$ to $x$ is because there is one and only one ordered pair in $f$ whose first component is $x$. Note that we aren't saying that $f$ is a function, we're just saying that $f$ is \emph{locally} a function on $x$ (it might well be a function in other points of its domain but we don't know that yet). Saying that in $f$ there is exactly one ordered pair whose first component is $x$ is the same than saying that there are no ordered pairs in \Verb+G+ whose first component is $x$. We say this by using the composition operator defined over binary relations, namely \Verb+comp+ (see Table \ref{t:reloper}): \Verb+comp({[X,X]},G,{})+. Indeed, this predicate says that when \Verb+{[X,X]}+ is composed with \Verb+G+ the result is the empty set. This can happen for two reasons: \Verb+G+ is the empty binary relation, in which case it's obvious that there are no ordered pairs with first component \Verb+X+; or \Verb+G+ is non-empty but no pair in it composes with \Verb+[X,X]+, which is equivalent to say that \Verb+X+ does not belong to the domain of \Verb+G+. We could have said the same by stating that \Verb+dom(G,D) & X nin D+ but this is usually less efficient because it requires to compute the domain of \Verb+G+.

Therefore, \verb+applyTo(F,X,Y)+ implies that \verb+X+ belongs to the domain of \verb+F+. If this is not the case then \verb+applyTo(F,X,Y)+ fails. Then, if we have to translate $x \in \dom f \land f(x) = y$ it's enough to state \verb+applyTo(F,X,Y)+. 

However, if in B we have that $f \in T \pfun U$ is part of the invariant, then $x \in \dom(f) \land f(x) = y$ will be defined due to the invariant. That is, $f(x)$ will be a unique value. This means that encoding it as \verb+applyTo(F,X,Y)+ is too much because \verb+applyTo+ asserts that \verb+F+ is locally a function on \verb+X+. Hence, in this case, a more precise encoding is the one based on set unification:
\begin{Verbatim}
F = {[X,Y] / G} & [X,Y] nin G
\end{Verbatim}
Note that this encoding implies that \verb+X+ belongs to the domain of \verb+F+ (otherwise it will fail as \verb+applyTo+). More importantly, this encoding is saying that all we have to do to find the image of \verb+X+ under \verb+F+ is to walk through \verb+F+ looking for \emph{the} ordered pair whose first component is \verb+X+. On the other hand, the encoding based on \verb+applyTo+ is saying that once we have found \verb+[X,Y]+ in \verb+F+ we have to keep walking through it to check that there's no other pair whose first component is \verb+X+. This last check required by \verb+applyTo+ is redundant if we know that \verb+F+ is a function. If we have proved that $f \in T \pfun U$ is an invariant then we know for sure that $f$ is a function.

Observe that in the translation of \textbf{findBirthday} we have used \Verb+applyTo+ which, after the above analysis, is not the best choice because  $pfun(birthday)$ is intended to be an invariant of the specification. We should replace \Verb+applyTo+ by the encoding based on set unification. We didn't do it in that way because we think that it requires a rather complex explanation when we were just introducing \setlog.

\subsection{\label{logicos}Translating logical operators}
Logical conjunction (\Verb+&+), disjunction (\Verb+or+), implication (\Verb+implies+) and negation (\Verb+neg+) are among the available logical connectives in \setlog (see Section 3.3 of the \setlog manual\footnote{\url{https://www.clpset.unipr.it/SETLOG/manual_4_9_8.pdf}} for the complete list). Logical negation (\Verb+neg+) must be used with care because, as the manual explains in Section 3.3, it doesn't work well in all cases. In general, \Verb+neg+ works as expected when the formula to be negated doesn't contain existential variables inside it. For instance, the following formula states that \Verb+Min+ is the minimum element in \Verb+S+:
\begin{Verbatim}
Min in S & subset(S,int(Min,Max))
\end{Verbatim}
\Verb+neg+ won't work correctly for this formula because \Verb+Max+ is an existential variable inside the formula. In order to see that \Verb+Max+ is an existential variable inside the formula, we can write it as the body of a clause computing the minimum element of a set:
\begin{Verbatim}
min(S,Min) :- Min in S & subset(S,int(Min,Max)).
\end{Verbatim}
Now it's clear that \Verb+Max+ is an existential variable inside the formula because it's not an argument of the clause head. Hence, \Verb+neg+ won't work well for \Verb+min+. More precisely, if we define the clause \Verb+n_min+ as follows:
\begin{Verbatim}
n_min(S,Min) :- neg(Min in S & subset(S,int(Min,Max))).
\end{Verbatim}
it doesn't correspond to $\lnot \texttt{min(S,Min)}$ because \Verb+neg+ won't compute the (correct) negation of its argument as it contains \Verb+Max+. \Verb+neg+ will compute some formula but not the negation we're expecting.

On the other hand, \setlog provides the negation for all its atomic constraints (Tables \ref{t:setoper}-\ref{t:reloper} and all the arithmetic constraints). \verb+neg+ works correctly for all of them. For example, if we want to translate $\lnot x \in A$ we can write in \setlog \Verb+neg(X in A)+ or just \Verb+X nin A+. In the same way, $\lnot A = b$ can be translated as \Verb+neg(A neq b)+ or as \Verb+A neq b+. For instance, the B predicate $A \not\subseteq B$ is translated as \Verb+nsubset(A,B)+; and $\lnot a \leq y$ as \Verb+neg(A =< Y)+. Tables \ref{t:setoper}-\ref{t:reloper} include the negation for every set theoretic operator.

As an example of using \Verb+neg+, the following B statement:
\[
\bite{x \in \dom(f) \land 0 < x}{f, msg := \{x\} \ndres f, ok}{msg := error}
\]
can be translated as follows:
\begin{Verbatim}
dom(F,D) &
(X in D & 0 < X & dares({X},F,F_) & Msg = ok
or
 neg(X in D & 0 < X) & Sa_ = Sa & Msg = error
)
\end{Verbatim}
Note that \Verb+dom(F,D)+ is placed outside the disjunction because the constraint is used to name the domain of \verb+F+. Observe that \verb+D+ isn't present in the B statement; it has to be introduced in \setlog to name the expression $\dom(f)$. \Verb+dom(F,D)+ states that \Verb+D+ is the (name of the) domain of \Verb+Sa+: it makes no sense to negate this because we're \emph{defining} \Verb+D+ as such. This situation arises frequently when a B specification is translated into \setlog due to the fact that B uses expressions for what in \setlog is written with predicates.

\subsubsection{\label{quatifiers}Quantifiers}
In general existential quantifiers need not to be translated because \setlog semantics is based on existentially quantifying all variables of any given program. For example, if in B we have:
\[
\exists x . (x \in \nat \land x \in A)
\]
it can be translated as:
\begin{Verbatim}
0 =< X & X in A
\end{Verbatim}
because the semantics of the \setlog program is, essentially, an existential quantifier over both variables.

Things are different when dealing with universal quantifiers. In \setlog we only have so-called \emph{restricted universal quantifiers} (RUQ). A RUQ is a formula of the following form:
\[
\forall x \in A : P(x)
\]
whose semantics is:
\[
\forall x . (x \in A \implies P(x))
\]
which, as can be seen, coincides with the universally quantified predicates available en B.

In \setlog the simplest RUQ are encoded as follows:
\begin{Verbatim}
foreach(X in A,P(X))
\end{Verbatim}
There are more complex and expressive RUQ available in \setlog\footnote{Have a look at chapter 6 of \setlog user's manual and then ask for help to the instructor.}.

Recall that a proper use of the B language tends to avoid most of the quantified formulas.

\subsection{Translating $\nat$}
As $\nat$ is not a type and, at the same time, is an interpreted set, we must be careful when translating $\nat$ into \setlog. 

A type declaration such as $x \in \nat$ is equivalent to $x \in \num \land 0 \leq x$. As we have said, $x \in \num$ is encoded in terms of the type system defined in \setlog, whereas $0 \leq x$ is simply encoded as \verb+0 =< X+. On the other hand, $A \subseteq \nat$ or $A \in \power \nat$ are translated with a RUQ:
\begin{Verbatim}
foreach(X in A, 0 =< X)
\end{Verbatim}
In particular a type declaration such as $f \in T \pfun \nat$ is encoded in \setlog as follows:
\begin{Verbatim}
pfun(F) & foreach([X,Y] in F, 0 =< Y)
\end{Verbatim}
plus a type declaration for $f$ such as \verb+dec(F,rel(t,int))+, assuming $T$ is a basic type.

\section{\label{simulacion}Running \setlog forgrams}

\setlog forgrams usually won't meet the typical performance requirements demanded by users. Forgrams are slower than programs but they have computational properties that programs don't. Hence, we see a \setlog forgram of a B specification more as a \emph{prototype} than as a final program. On the other hand, given the similarities between a B specification and the corresponding \setlog forgram, it's reasonable to think that the prototype is a \emph{correct} implementation of the specification\footnote{In fact, the translation process can be automated in many cases.}. Then, we can use these prototypes to make an early validation of the requirements. 

Validating user requirements by means of prototypes entails executing the prototypes together with the users so they can agree or disagree with the behavior of the prototypes. This early validation will detect many errors, ambiguities and incompleteness present in the requirements and possible misunderstandings or misinterpretations caused by the software engineers. Without this validation many of these issues would be detected in later stages of the project thus increasing the project costs. Think that if one of these issues is detected once the product has been put in the market, it implies to correct the error in the requirements document, the specification, the design, the implementation, the user documentation, etc.

Since we see \setlog forgrams as prototypes we talk about \emph{simulations} or \emph{animations} rather than \emph{executions} when speaking about running them. However, technically, what we do is no more than executing a piece of code. The word \emph{simulation} is usually used in the context of \emph{models} (e.g. modeling and simulation). In a sense, our \setlog forgrams are \emph{executable models} of the user requirements. On the other hand, the word \emph{animation} is usually used in the context of formal specifications. In this sense, the \setlog implementation of a B specification can be seen as an \emph{executable specification}. In fact, as we will see, \setlog forgrams have features and properties usually enjoyed by specifications and models, which are rare or nonexistent in programs written in imperative (and even functional) programming languages.

Be it execution, simulation or animation the basic idea is to provide inputs to the forgram, model or specification and observe the produced outputs or effects. Besides, we will show that \setlog offers more possibilities beyond this basic idea.

\subsection{\label{simples}Basic simulations}
Let's see an example of a simulation on a \setlog forgram. Assume the forgram of the birthday book is saved in a file named \Verb+bb.pl+. We start by executing the Prolog interpreter from a command line terminal and from the folder where \setlog was installed\footnote{The name of the Prolog executable may vary depending on the interpreter and the operating system. The example corresponds to a Ubuntu Linux machine and SWI-Prolog.}.
\begin{verbatim}
~/setlog$ prolog

?- consult('setlog.pl').

?- setlog.

{log}=> consult('bb.pl').

{log}=> birthdayBookInit(K,B) & addBirthday(K,B,maxi,160367,K_,B_,M).
K = {},  
B = {},  
K_ = {maxi},  
B_ = {[maxi,160367]},  
M = ok

Another solution?  (y/n) y
no
{log}=> 
\end{verbatim}
The meaning of the above code is the following:
\begin{enumerate}
\item The Prolog interpreter is executed.
\item The \setlog interpreter is loaded.
\item The \setlog interpreter is accessed.
\item The birthday book prototype is loaded.
\item The simulation is run:
\begin{Verbatim}
birthdayBookInit(K,B) & addBirthday(K,B,maxi,160367,K_,B_,M).
\end{Verbatim}
consisting of:
\begin{itemize}
\item \Verb+birthdayBookInit+ is called passing to it any two variables as arguments;
\item \Verb+addBirthday+ is called passing to it in the first and second arguments the same variables used to call \Verb+birthdayBookInit+; as the third and fourth arguments two constants; and three new variables in the last three arguments.
\end{itemize}
Observe that the simulation ends in a dot.
\item \setlog shows the result of the simulation.
\item \setlog asks if we want to see other solutions and we answer yes.
\item \setlog says there are no more solutions.
\end{enumerate}

Let's see the simulation in detail:
\begin{Verbatim}
birthdayBookInit(K,B) & addBirthday(K,B,maxi,160367,K_,B_,M).
\end{Verbatim}
When we call \Verb+birthdayBookInit(K,B)+, \Verb+K+ and \Verb+B+ unify with  \Verb+Known+ and \Verb+Birthday+ which are the formal arguments used in the definition of \Verb+birthdayBookInit+ (see the complete code in Appendix \ref{ap:bb}). This implies that \Verb+K+ is equal to \Verb+Known+ and \Verb+B+ is equal to \Verb+Birthday+ which in turn implies that \Verb+K+ and \Verb+B+ are equal to \Verb+{}+. This is exactly the first line of the answer returned by \setlog. Hence, when \Verb+addBirthday(K,B,maxi,160367,K_,B_,M)+ is called, it's like we were calling:
\begin{Verbatim}
addBirthday({},{},maxi,160367,K_,B_,M)
\end{Verbatim}
Calling \Verb+addBirthday+ makes \setlog to execute each branch of the disjunction present in the body of the clause. That is, both branches are  tried in the order they're written. Then, unification goes as follows:
\begin{Verbatim}
Known = {}
Birthday = {}
Name_i = maxi
Date_i = 160367
K_ = Known_
B_ = Birthday_
M = Msg
\end{Verbatim}
Hence the code in the first branch is instantiated as follows:
\begin{Verbatim}
maxi nin {} &
un({},{maxi},K_) &
un({},{[maxi,160367]},B_) &
M = ok
\end{Verbatim}
which reduces to:
\begin{Verbatim}
K_ = {maxi} &
B_ = {[maxi,160367]} &
M = ok
\end{Verbatim}
which corresponds to the second line of the answer returned by \setlog.

When '\Verb+y+' is pressed \setlog executes the second branch. Again, unification takes place and a new series of equations are produced:
\begin{Verbatim}
Known = {}
Birthday = {}
Name = maxi
K_ = Known
B_ = Birthday
M = Msg
\end{Verbatim}
which implies that \Verb+K+ unifies with \Verb+{}+. Then, the code in the second branch is instantiated as follows:
\begin{Verbatim}
maxi in {} ...
\end{Verbatim}
As this predicate is obviously false, the invocation of this branch fails and hence \setlog produces no solution. As a consequence \setlog answers \Verb+no+ after we press '\Verb+y+'.

The following simulation is longer and includes the previous one.
{\small
\begin{Verbatim}
birthdayBookInit(K,B)                     & addBirthday(K,B,maxi,160367,K1,B1,M1) &
addBirthday(K1,B1,'Yo',201166,K2,B2,M2)   & findBirthday(K2,B2,'Yo',C,K3,B3) &
addBirthday(K3,B3,'Otro',201166,K4,B4,M4) & remind(K4,B4,160367,Card,K5,B5) &
remind(K5,B5,201166,Card1,K_,B_).
\end{Verbatim}
}
\noindent Here we can see that we're calling all the operations defined in the prototype; that we use different variables to chain the state transitions; and that it's possible to use constants beginning with an uppercase letter as long as we enclose them between single quotation marks.
The first solution returned by that simulation is the following:
\begin{Verbatim}
K = {},  
B = {},  
K1 = {maxi},  
B1 = {[maxi,160367]},  
M1 = ok,  
K2 = {maxi,Yo},  
B2 = {[maxi,160367],[Yo,201166]},  
M2 = ok,  
C = 201166,  
K3 = {maxi,Yo},  
B3 = {[maxi,160367],[Yo,201166]},  
K4 = {maxi,Yo,Otro},  
B4 = {[maxi,160367],[Yo,201166],[Otro,201166]},  
M4 = ok,  
Card = {maxi},  
K5 = {maxi,Yo,Otro},  
B5 = {[maxi,160367],[Yo,201166],[Otro,201166]},  
Card1 = {Yo,Otro},  
K_ = {maxi,Yo,Otro},  
B_ = {[maxi,160367],[Yo,201166],[Otro,201166]}
\end{Verbatim}
where we can see that \setlog gives us the chance to have a complete trace of the forgram execution. Note also that \setlog eliminates the single quotation marks we used to enclose some constants.

It's important to remark that the variables used to chain the state transitions (i.e. \Verb+K1+, \Verb+B1+, \dots, \Verb+K5+, \Verb+B5+) must be all different. If done otherwise, the simulation might be incorrect. For instance:
\begin{Verbatim}
birthdayBookInit(K,B) & addBirthday(K,B,N,C,K,B,M).
\end{Verbatim}
will fail as the values of \Verb+K+ and \Verb+B+ before invoking \Verb+addBirthday+ can't unify with the values returned by it. In other words, the \Verb+K+ and \Verb+B+ as the first two arguments of \Verb+addBirthday+ can't have the same value than the \Verb+K+ and \Verb+B+ used towards the end of the call. We could use the same variable for the before and after state of query state operations (for instance when we invoke \Verb+findBirthday+ and \Verb+remid+).

So far the two simulations we have performed start in the initial state. It's quite simple to start a simulation from any state:
\begin{Verbatim}
K = {maxi,caro,cami,alvaro} &
B = {[maxi,160367],[caro,201166],[cami,290697],[alvaro,110400]} &
addBirthday(K,B,'Yo',160367,K1,B1,M1) & remind(K1,B1,160367,Card,K1,B1).
\end{Verbatim}
where we can see that we use the same variable to indicate the before and after state of \Verb+remid+ (because we know this clause produces no state change). In this case the answer is:
\begin{Verbatim}
K = {maxi,caro,cami,alvaro},  
B = {[maxi,160367],[caro,201166],[cami,290697],[alvaro,110400]},  
K1 = {maxi,caro,cami,alvaro,Yo},  
B1 = {[maxi,160367],[caro,201166],[cami,290697],[alvaro,110400],[Yo,160367]},  
M1 = ok,  
Card = {maxi,Yo}
\end{Verbatim}

A potential problem of manually defining the initial state for a simulation is that this state, due to human error, might not verify the state invariant. Nevertheless, it's very easy to avoid this problem as we will see in Section \ref{evaluacion}.

\subsubsection{Hiding the complete trace of the execution}
If we don't need the complete execution trace of a simulation but only the its final state and outputs we can define a clause for the simulation whose arguments are the variables we are interested in:
\begin{Verbatim}
sim(K_,B_,C,Card,Card1) :-
  birthdayBookInit(K,B) &
  addBirthday(K,B,maxi,160367,K1,B1,M1) &
  addBirthday(K1,B1,'Yo',201166,K2,B2,_) &
  findBirthday(K2,B2,'Yo',C,K3,B3) &
  addBirthday(K3,B3,'Otro',201166,K4,B4,_) &
  remind(K4,B4,160367,Card,K5,B5) &
  remind(K5,B5,201166,Card1,K_,B_).
\end{Verbatim}
And then we call the clause:
\begin{Verbatim}
{log}=> sim(K_,B_,C,Card,Card1).
K_ = {maxi,Yo,Otro},  
B_ = {[maxi,160367],[Yo,201166],[Otro,201166]},  
C = 201166,  
Card = {maxi},  
Card1 = {Yo,Otro}
\end{Verbatim}
\noindent As can be seen, we get a more compact output showing only the variables we are interested in.

\subsection{Type checking and simulations}
So far we haven't really used \setlog's typechecker. Actually when we consulted \Verb+bb.pl+ the types weren't checked. In other words \setlog ignored the \Verb+dec_p_type+ assertions included in \Verb+bb.pl+. This means that possible type errors weren't detected by \setlog. In this sense \setlog executed all the simulations in untyped mode. In this section we'll see how to call the typechecker and how this affects simulations. Recall reading chapter 9 of \setlog user's manual for further details on \setlog's types.

Type checking can be activated by means of the \Verb+type_check+ command which should be issued before the file is consulted.
\begin{verbatim}
~/setlog$ prolog

?- consult('setlog.pl').

?- setlog.

{log}=> type_check.         % typechecker is active

{log}=> consult('bb.pl').
\end{verbatim}
In this way, when \setlog executes \Verb+consult+ it invokes the typechecker and if there are type errors we'll see an error message.

Type checking can be deactivated at any time by means of command \Verb+notype_check+.

When the typechecker is active all simulations must be correctly typed because otherwise \setlog will just print a type error.
\begin{Verbatim}
{log}=> birthdayBookInit(K,B) & addBirthday(K,B,maxi,160367,K_,B_,M).

***ERROR***: type error: variable K has no type declaration
\end{Verbatim}
Then, we have to declare the type of all variables:
\begin{Verbatim}
{log}=> birthdayBookInit(K,B) & addBirthday(K,B,name:maxi,date:160367,K_,B_,M) & 
        dec([K,K_],kn) & dec([B,B_],bb) & dec(M,msg).

K = {},  
B = {},  
K_ = {name:maxi},  
B_ = {[name:maxi,date:160367]},  
M = ok
\end{Verbatim}

If the user wants to typecheck the program, for instance \Verb+bb.pl+, but (s)he doesn't want to deal with types when running simulations, the typechecker can be deactivated right after consulting the program. In this way \setlog will check the types of the program but it then will accept untyped simulations.

Clearly, in general, working with untyped simulations is easier but more dangerous because we could call the program with ill-typed inputs thus causing false failures.

In the rest of this section we'll work with untyped simulations. This means that the user must ensure that typechecking is deactivated (command \Verb+notype_check+).

\subsection{Simulations using integer numbers}
As we have said, \setlog is, essentially, a set solver. However, it's also capable of solving formulas containing predicates over the integer numbers. In that regard, \setlog uses two external solvers known as CLP(FD)\footnote{\url{https://www.swi-prolog.org/pldoc/man?section=clpfd-predicate-index}} and CLP(Q)\footnote{\url{https://www.swi-prolog.org/pldoc/man?section=clpqr}}. Each of them has its advantages and disadvantages.

By default \setlog uses CLP(Q). Users can change to CLP(FD) by means of command \Verb+int_solv+-\Verb+er(clpfd)+ and can come back to CLP(Q) by means of \Verb+int_solver(clpq)+.

Generally speaking, it's more convenient to run simulations when  CLP(FD) is active because it tends to generate more concrete solutions. In particular CLP(FD) is capable of performing labeling over the integer numbers which allows users to go through the solutions interactively. Labeling works if at least some of the integer variables are bound to a finite domain. Variable \Verb+N+ is bound to the finite domain \Verb+int(a,b)+ (\verb+a+ and \Verb+b+ integer numbers) if  \Verb+N in int(a,b)+ is in the formula. See chapter 7 of \setlog user's manual for more details.

For example, if CLP(Q) is active, the answer to the following goal:
\begin{verbatim}
Turn is 2*N + 1.
\end{verbatim}
is exactly the same formula. That is, \setlog is telling us that the formula is satisfiable but we don't have one of its solutions. If we activate CLP(FD):
\begin{verbatim}
int_solver(clpfd).

Turn is 2*N + 1.
\end{verbatim}
\setlog prints a warning message and the same formula:
\begin{verbatim}
***WARNING***: non-finite domain

true
Constraint: Turn is 2*N+1
\end{verbatim}
This means that the formula \emph{might be} satisfiable but  CLP(FD) isn't sure. If we want a more reliable answer we have to bound \Verb+Turn+ or \Verb+N+ to a finite domain:
\begin{verbatim}
N in int(1,5) & Turn is 2*N + 1.
\end{verbatim}
in which case the first solution is:
\begin{verbatim}
N = 1, Turn = 3
\end{verbatim}
and we can get more solutions interactively. On the contrary, if we activate CLP(Q) the finite domain doesn't quite help to get a concrete solution:
\begin{verbatim}
int_solver(clpq).

N in int(1,5) & Turn is 2*N + 1.

true
Constraint: N>=1, N=<5, Turn is 2*N+1
\end{verbatim}

On the other hand, CLP(Q) is complete for linear integer arithmetic while CLP(FD) isn't. This means that if we want to use \setlog to \emph{automatically prove} a property of the program \emph{for all the integer numbers}, we must use CLP(Q)\footnote{As in general non-linear arithmetic is undecidable it's quite difficult to build a tool capable of automatically proving program properties involving non-linear arithmetic.}. Given that simulations don't prove properties it's reasonable to use CLP(FD).

\subsection{Symbolic simulations}
The symbolic execution of a program means to execute it providing to it variables as inputs instead of constants. This means that the execution engine should be able to symbolically operate with variables in order to compute program states as the execution moves forward. As a symbolic execution operates with variables, it can show more general properties of the program than when this is run with constants as input.

\setlog is able to symbolically execute forgrams, within certain limits. These limits are given by set theory and non-recursive clauses. The following are the conditions under which \setlog can perform symbolic executions\footnote{This is an informal description and not entirely accurate of the conditions for \setlog being able to perform symbolic executions. These conditions are more or less complex and quite technical. The \setlog forgrams that can't be symbolically simulated and don't verify the following conditions will not appear in this course.}:
\begin{enumerate}
  \item Recursive clauses are not allowed.
  \item Only the operators of Tables \ref{t:setoper} and \ref{t:reloper} are allowed. If the \setlog forgrams uses the cardinality operator (\Verb+size+), the program can't use the operators of Table \ref{t:reloper}. The \Verb+size+ operator is complete only when combined with the operators of Table \ref{t:setoper}.
  \item All the arithmetic formulas are linear\footnote{More precisely, all the integer expressions must be sums or subtractions of terms of the form \Verb+x*y+ with \Verb+x+ or \Verb+y+ constants. All arithmetic relational operators are allowed, even \Verb+neq+.}. 
\end{enumerate}
This means the \setlog code can't use operators of Table \ref{t:listoper} if symbolic executions are to be done\footnote{The problem with the operators of Table \ref{t:listoper} is that they depend on certain aspects of set theory that aren't fully implemented in \setlog, yet.}. Actually, many symbolic executions are still possible even if the above conditions aren't met.

The \setlog forgram of the birthday book falls within the limits of what \setlog can symbolically execute. For example, starting from the initial state we can call \Verb+addBirthday+ using just variables:
\begin{Verbatim}
birthdayBookInit(K,B) & addBirthday(K,B,N,C,K_,B_,M).
\end{Verbatim}
in which case \setlog answers:
\begin{Verbatim}
K = {},  
B = {},  
K_ = {N},  
B_ = {[N,C]},  
M = ok
\end{Verbatim}
which is a representation of the expected results. Now we can chain a second invocation to \Verb+addBirthday+ using other input variables:
\begin{Verbatim}
birthdayBookInit(K,B) & 
addBirthday(K,B,N1,C1,K1,B1,M1) & addBirthday(K1,B1,N2,C2,K_,B_,M2).
\end{Verbatim}
in which case the first solution returned by \setlog is:
\begin{Verbatim}
K = {},  
B = {},  
K1 = {N1},  
B1 = {[N1,C1]},  
M1 = ok,  
K_ = {N1,N2},  
B_ = {[N1,C1],[N2,C2]},  
M2 = ok
Constraint: N1 neq N2
\end{Verbatim}

As can be seen, the answer includes the \verb+Constraint+ section which has never appeared before. Indeed, the most general solution that can be returned by \setlog consists of two parts: a (possibly empty) list of equalities between variables and terms (or expressions); and a (possibly empty) list of \emph{constraints}. Each constraint is a \setlog predicate; the returned constraints appear after the word \Verb+Constraint+. The conjunction of all these constraints is always satisfiable (in general the solution is obtained by substituting the variables of type set by the empty set). In this example, clearly, the second invocation to \Verb+addBirthday+ can add the pair \Verb+[N2,C2]+ to the birthday book if and only if \Verb+N2 nin {N1}+, which holds if and only if \Verb+N2+ is different from \Verb+N1+.

\setlog returns a second solution to this symbolic execution:
\begin{Verbatim}
K = {},  
B = {},  
K1 = {N1},  
B1 = {[N1,C1]},  
M1 = ok,  
N2 = N1,  
K_ = {N1},  
B_ = {[N1,C1]},  
M2 = nameExists
\end{Verbatim}
produced after considering that \Verb+N1+ and \Verb+N2+ are equal in which case the second invocation to \Verb+addBirthday+ goes through the \belsep branch and so \Verb+K_+ and \Verb+B_+ are equal to \Verb+K1+ and \Verb+B1+, which is the expected result as well.

Clearly, symbolic executions allows us to draw more general conclusions about the behavior of the prototype. The next example illustrates this:
\begin{Verbatim}
birthdayBookInit(K,B) & addBirthday(K,B,N1,C1,K1,B1,M1) &
addBirthday(K1,B1,N2,C2,K2,B2,M2) & findBirthday(K2,B2,W,X,K2,B2).
\end{Verbatim}
\setlog will consider several particular cases depending on whether \Verb+N2+, \Verb+N1+ and \Verb+W+ are equal or not. For example, the following are the first three solutions returned by \setlog:
\begin{Verbatim}
K = {},  
B = {},  
K1 = {N1},  
B1 = {[N1,C1]},  
M1 = ok,  
K2 = {N1,N2},  
B2 = {[N1,C1],[N2,C2]},  
M2 = ok,  
W = N1,  
X = C1
Constraint: N1 neq N2

Another solution?  (y/n)
K = {},  
B = {},  
K1 = {N1},  
B1 = {[N1,C1]},  
M1 = ok,  
K2 = {N1,N2},  
B2 = {[N1,C1],[N2,C2]},  
M2 = ok,  
W = N1,  
X = C1
Constraint: C1 neq C2, N1 neq N2

Another solution?  (y/n)
K = {},  
B = {},  
K1 = {N1},  
B1 = {[N1,C1]},  
M1 = ok,  
K2 = {N1,N2},  
B2 = {[N1,C1],[N2,C2]},  
M2 = ok,  
W = N2,  
X = C2
Constraint: N1 neq N2
\end{Verbatim}
In the first case \Verb+W = N1+ is considered and so \Verb+X+ must be equal to \Verb+C1+; the second case is similar to the first one; and in the third \Verb+W = N2+ and so \Verb+X+ is equal to \verb+C2+. \setlog returns more solutions some of which are repeated.

Obviously symbolic simulations may combine variables with constants. In general the less the variables we use the less the number of solutions.

\subsection{Inverse simulations}
Normally, in a simulation the user provides inputs and the forgram returns the outputs. There are situations in which is interesting to get the inputs from the outputs. This means a sort of an inverse simulation.

\setlog is able to perform inverse executions within the same limits in which it is able to perform symbolic executions. In fact, a careful reading of the previous section reveals that \setlog doesn't really distinguish input from output variables, nor between before and after states. As a consequence, for \setlog is more or less the same to execute a forgram by providing values for the input variables or for the output variables; in fact, \setlog is able to execute a forgram just with variables.

Let's see a very simple inverse simulation where we only give the after state:
\begin{Verbatim}
K_ = {maxi,caro,cami,alvaro} &
B_ = {[maxi,160367],[caro,201166],[cami,290697],[alvaro,110400]} &
addBirthday(K,B,N,C,K_,B_,M).
\end{Verbatim}
The first solution returned by \setlog is the following:
\begin{Verbatim}
K_ = {maxi,caro,cami,alvaro},  
B_ = {[maxi,160367],[caro,201166],[cami,290697],[alvaro,110400]},  
K = {maxi,caro,cami},  
B = {[maxi,160367],[caro,201166],[cami,290697]},  
N = alvaro,  
C = 110400,  
M = ok
\end{Verbatim}

When the B specification is deterministic, the corresponding \setlog forgram will be deterministic as well. Therefore, for any given input there will be only one solution. However, the inverse simulation of a deterministic forgram may generate a number of solutions. This is the case with the above simulation. The first solution computed by \setlog considers the case where \Verb+N = alvaro+ and \Verb+C = 110400+, but this isn't the only possibility. Going forwards with the solutions we get, for instance, the following:
\begin{Verbatim}
K_ = {maxi,caro,cami,alvaro},  
B_ = {[maxi,160367],[caro,201166],[cami,290697],[alvaro,110400]},  
K = {maxi,caro,alvaro},  
B = {[maxi,160367],[caro,201166],[alvaro,110400]},  
N = cami,  
C = 290697,  
M = ok
\end{Verbatim}
which means that \Verb+K_+ and \Verb+B_+ may have been generated by starting from some \Verb+K+ and \Verb+B+ where \Verb+cami+'s birthday isn't in the book and so we can add it.

\subsection{\label{evaluacion}Evaluation of predicates}
At the end of Section \ref{simples} we showed how to start a simulation from a state different from the initial state. We also said that this entails some risks as manually writing the start state is error prone which may lead to an unsound state. In this section we will see how to avoid this problem by using a feature of \setlog that is useful for other verification activities, too.

Let's consider the following state of the birthday book:
\begin{Verbatim}
Known = {maxi,caro,cami,alvaro}
Birthday = {[maxi,160367],[caro,201166],[cami,290697],[alvaro,110400]}
\end{Verbatim}
Starting a simulation from this state may give incorrect results if it doesn't verify the state invariant defined for the specification. Recall that the state invariant for the birthday book is \verb+birthdayBookInv(Known,Birthday)+.

Hence, we can check whether or not the above state satisfies the invariant by asking \setlog to solve the following:
\begin{Verbatim}
Known = {maxi,caro,cami,alvaro} &
Birthday = {[maxi,160367],[caro,201166],[cami,290697],[alvaro,110400]} &
birthdayBookInv(Known,Birthday).
\end{Verbatim}
in which case \setlog returns the values of \Verb+Known+ and \Verb+Birthday+, meaning that \Verb+birthdayBookInv+ is satisfied. If this weren't the case the answer would have been \Verb+no+, as in the following example (note that \Verb+maxi+ is missing from \Verb+known+):
\begin{Verbatim}
Known = {caro,cami,alvaro} &
Birthday = {[maxi,160367],[caro,201166],[cami,290697],[alvaro,110400]} &
birthdayBookInv(Known,Birthday).
\end{Verbatim}

\section{\label{pruebas}Proving the correctness of \setlog forgrams}
Evaluating properties with \setlog helps to run correct simulations by checking that the starting state is correctly defined. It also helps to \emph{test} whether or not certain properties are true of the specification or not. However, it would be better if we could  \emph{prove} that these properties are true of the specification. In this section we will see how \setlog allows us to prove that the operations of a specification preserve the state invariant.

So far we have used \setlog as a programming language. However, \setlog is also a \emph{satisfiability solver}\footnote{See for instance Wikipedia: \href{https://en.wikipedia.org/wiki/Satisfiability_modulo_theories}{Satisfiability modulo theories}.}. This means that \setlog is a program that can decide if formulas of some theory are \emph{satisfiable} or not. In this case the theory is the theory of finite sets and binary relations given by the operators listed in Tables \ref{t:setoper} and \ref{t:reloper}, and combined with linear integer arithmetic\footnote{In what follows we will only mention the theory of finite sets but the same is valid for this theory combined with linear integer algebra.}. 

If $F$ is a formula depending on a variable, we say that $F$ is \emph{satisfiable} if and only if:
\[
\exists y: F(y)
\]

In the case of \setlog, $y$ is quantified over \emph{all} finite sets. Therefore, if \setlog answers that $F$ is satisfiable it means that there exists a finite set satisfying it. Symmetrically, if \setlog says that $F$ is unsatisfiable it means that there is no finite set satisfying it. Formally, $F$ is an unsatisfiable formula if:
\begin{equation}\label{eq:unsat}
\forall y: \lnot F(y)
\end{equation}
where $y$ ranges over all finite sets. If we call $G(x) \defs \lnot F(x)$ then \eqref{eq:unsat} becomes:
\begin{equation}\label{eq:sat}
\forall y: G(y)  
\end{equation}
which means that $G$ is true of every finite set. Putting it in another way, $G$ is \emph{valid} with respect to the theory of finite sets; or, equivalently, $G$ is a \emph{theorem} of the theory of finite sets.

\begin{tcolorbox}[%
  toprule=2mm,
  before skip=10pt plus 2pt,
  after skip=10pt plus 2pt]
In summary, if \setlog decides that \emph{$F$ is unsatisfiable}, then we know that \emph{$\lnot F$ is a theorem}.
\end{tcolorbox}

In other words, \eqref{eq:unsat} and \eqref{eq:sat} are two sides of the same coin: \eqref{eq:unsat} says that $F$ is unsatisfiable and \eqref{eq:sat} says that $G$ (i.e. $\lnot F$) is a theorem.

If \setlog is called on some formula there are four possible behaviors:
\begin{enumerate}
\item \setlog returns \verb+no+. This means the formula is unsatisfiable.
\item \setlog returns one or more solutions. This means the formula is satisfiable. For example, the simulations we run in Section \ref{simulacion} are all satisfiable formulas.
\item \setlog returns a warning messages. This means the answer is unreliable. We can't be sure whether the formula is satisfiable or not.
\item \setlog doesn't seem to return. You wait in front of the screen after pressing the return key but no answer is produced; you wait longer but still nothing happens. This means that \setlog is unable to determine whether the formula is satisfiable or not. This in turn may occur because the formula is too complex and makes \setlog to take a very long time of just because \setlog enters into an infinite loop. Situations like this are rare and usually occur in complex problems. If you want to see this behavior try the following:
\begin{Verbatim}
comp(R,R,R) & [X,Y] in R & [Y,Z] in R & [X,Z] nin R.
\end{Verbatim}
What is the meaning of this formula?
\end{enumerate}

One important aspect is that \setlog, as other satisfiability solvers, \emph{automatically} decides the satisfiability of a given formula. That is, no action from the user is required. Hence, when \setlog finds that $F$ is unsatisfiable it has \emph{automatically proved} the theorem $\lnot F$. This is called \emph{automated theorem proving} which is part of \emph{automated software verification}. There are, however, automated theorem provers that aren't satisfiability solvers\footnote{See for instance Wikipedia: \href{https://en.wikipedia.org/wiki/Automated_theorem_proving}{Automated theorem proving}.}. Satisfiability solvers and automated theorem provers can be used to prove mathematical theorems but we're interested in their application to software verification. 

More specifically, we're going to apply \setlog's capabilities for automated theorem proving to ensure machine consistency. Recall that in Section 5 of ``Introduction to the B-Method'' we show that the B-Method requires to discharge some proof obligations once we have written a B machine. Then, we're going to use \setlog to discharge those proof obligations on the corresponding \setlog forgram. That is, once we have translated the B specification into \setlog, we're going to use \setlog to generate the same proof obligations required by the B-Method and then we're going to use \setlog again to \emph{automatically} discharge them. This process implies that the forgram so verified becomes a \emph{certified prototype} of the system. In other words, the forgram is an implementation verifying all the the verification conditions set forth by the B-Method.

\subsection{Invariance lemmas in \setlog}

The most complex verification conditions required by the B-Methods are the invariance lemmas. 
Recall that an invariance lemma states that each operation of a B specification preserves the state invariant. Formally, if an operation depends on an input parameter $x$, has precondition $Pre$ and changes state variable $v$ with $Post$, the invariance lemma is as follows:
\begin{gather*}
\forall x. (Inv \land Pre \implies Inv[v \mapsto Post])
\end{gather*}
In turn, when this operation is translated as a \setlog clause we have $v\_$ as the next-state variable. The abstract assignment $v := Post$ becomes an equality of the form $v\_ = Post$. Therefore, the invariance lemma can be written as follows:
\begin{gather*}
\forall x. (Inv \land Pre \implies Inv[v \mapsto v\_])
\end{gather*}
If we define $Inv\_$ as a shorthand for $Inv[v \mapsto v\_]$, then we have:
\begin{gather*}
\forall x. (Inv \land Pre \implies Inv\_)
\end{gather*}
Recall that in order to prove the above formula in \setlog we must negate it:
\begin{gather*}
\lnot(\forall x. (Inv \land Pre \implies Inv\_))
\end{gather*}

At the same time during the translation of the B-Machine into \setlog, we have split the invariance in several pieces. Recall that for the birthday book specification we have the following:
\begin{Verbatim}
birthdayBookInv(Known,Birthday) :- dom(Birthday,Known) & pfun(Birthday).
\end{Verbatim}
Then, for instance, this is the invariance lemma for \verb+addBirthday+:
\begin{Verbatim}
addBirthday_pi_birthdayBookInv :-
  neg(
    birthdayBookInv(Known,Birthday) &
    addBirthday(Known,Birthday,Name,Date,Known_,Birthday_,Msg) implies
    birthdayBookInv(Known_,Birthday_)
  ).
\end{Verbatim}
The idea is that the user executes \verb+addBirthday_pi_birthdayBookInv+ and \setlog answers \verb+no+. As we have said above, this means that \setlog couldn't find values for the variables as to satisfy the formula (i.e. the formula is unsatisfiable). In turn, as we have explained, this means that the formula inside \verb+neg+ is a theorem and so \setlog has discharged this proof obligation.

There's, though, a problem that we need to address. Internally, \setlog transforms the body of \verb+addBirthday_pi_birthdayBookInv+ in:
\begin{Verbatim}
birthdayBookInv(Known,Birthday) &
addBirthday(Known,Birthday,Name,Date,Known_,Birthday_,Msg) &
neg( birthdayBookInv(Known_,Birthday_) ).
\end{Verbatim}
because $\lnot(I \land T \implies I\_) \equiv \lnot(\lnot(I \land T) \lor I\_) \equiv I \land T \land \lnot I\_$. The problem is that \setlog can't compute the negation of user-defined clauses. Then, \setlog will issue a warning such as:
\begin{Verbatim}
***WARNING***: Unsafe use of negation - using naf
\end{Verbatim}
In order to avoid this problem we have to help \setlog to compute the negation of the clauses declared as invariants. More precisely, we have to add the following to the birthday book forgram:
\begin{Verbatim}
dec_p_type(n_birthdayBookInv(kn,bb)).
n_birthdayBookInv(Known,Birthday) :- neg(dom(Birthday,Known) & pfun(Birthday)).
\end{Verbatim}
That is, for each clause \verb+p+ declared as an invariant, a clause named \verb+n_p+ with the same arity and whose body is the negation of \verb+p+'s body, is added to the forgram. In this way when \setlog has to compute \verb+neg(birthdayBookInv(Known_,Birthday_))+ it looks up among the clauses one whose head is \verb+n_birthdayBookInv+ and with \verb+birthdayBookInv+'s arity. If such a clause is present, \setlog uses its body to compute the negation; otherwise it issues a warning message such as the one above. These clauses are called \emph{negative clauses}. Note that negative clauses aren't declared as invariants although their types are those of the corresponding positive clauses. See Appendix \ref{ap:bb} for the complete forgram implementing the birthday book.

Recall that \Verb+neg+ doesn't always work correctly, as we explained in Section \ref{logicos}. However, it works well in many cases. You won't see problems with \verb+neg+ in what concerns the exercises of this course. You can have a look at the problem of computing $\lnot p$ in logic programming in Wikipedia: \href{https://en.wikipedia.org/wiki/Negation_as_failure}{Negation as failure}.

In any case, if you are in front of a formula for which \verb+neg+ doesn't work well, you can manually write its negation and put it in a negative clause. To that end you have to distribute the negation all the way down to the atoms at which point you use the negations of the operators of Tables \ref{t:setoper} and \ref{t:reloper}.

\subsection{The verification condition generator (VCG)}
\setlog can automatically generate verification conditions similar to those required by the B-Method, plus some more not required by the B-Method. That is, \setlog generates verification conditions as those discussed in Section 5 of ``Introduction to the B-Method''. We'll exemplify the process to generate verification conditions with the birthday book forgram.
\begin{verbatim}
~/setlog$ prolog

?- consult('setlog.pl').

?- setlog.

{log}=> vcg('bb.pl').
\end{verbatim}

VGC stands for \emph{verification condition generator}. The command takes as argument the name of a file containing a forgram implementing a state machine (in particular one resulting from the translation of a B machine). That is, the forgram must have declarations such as \verb+variables+, \verb+invariant+, etc. as described in Section \ref{traduccion} and in chapter 11 of the \setlog user's manual. VCG checks some well-formedness conditions on the forgram as described in detail in the referred manual. If all these checks are passed then VCG generates a file named, for instance, \verb+bb-vcg.pl+. Appendix \ref{ap:bb-vcg} lists the contents of \verb+bb-vcg.pl+ as produced by VCG. 

Once VCG has been called on a file, the user has to consult the file generated by VCG and run the command indicated by \setlog:
\begin{verbatim}
{log}=> consult('bb-vcg.pl').

Type checking has been deactivated.

Call check_vcs_bb to run the verification conditions.

file bb-vc.pl consulted.
\end{verbatim}
As can be seen, \setlog says that we should call \verb+check_vcs_bb+ to run or discharge the verification conditions. This command is always of the form \verb+check_vcs_<filename>+. If we run the command we'll see the following:
\begin{Verbatim}
{log}=> check_vcs_bb.

Checking birthdayBookInit_sat_birthdayBookInv ... OK
Checking addBirthday_is_sat ... OK
Checking findBirthday_is_sat ... OK
Checking remind_is_sat ... OK
Checking addBirthday_pi_birthdayBookInv ... OK
Checking findBirthday_pi_birthdayBookInv ... OK
Checking remind_pi_birthdayBookInv ... OK
\end{Verbatim}

As you can see, \setlog is able to automatically discharge all proof obligations. However, this might not always be the case. Why \setlog might be unable to discharge a proof obligation and how to remedy this situation is explained in the next section.

VCG generates basically two classes of verification conditions:
\begin{itemize}
\item \textsc{Satisfiability Conditions}. These are identified by the word \verb+_sat_+. For example, \verb+addBirth+-\verb+day_is_sat+ and \verb+birthdayBookInit_sat_birthdayBookInv+.

The expected answer for a satisfiability condition is a solution. In other words, if \setlog answers \verb+no+ for such a verification condition there's an error in the specification.
\item \textsc{Invariance Lemmas}. These are identified by the word \verb+_pi_+ (for ``preserves invariant''). For example, \verb+addBirthday_pi_birthdayBookInv+.

The expected answer for an invariance lemma is \verb+no+. In other words, if \setlog returns a solution for such a verification condition there's an error in the specification.
\end{itemize}

\subsection{When \setlog fails to discharge a proof obligation}
We'll focus this section on invariance lemmas but similar conclusions can be drawn for satisfiability conditions. \setlog may fail to discharge (i.e. prove) an invariance lemma, basically, for two reasons:
\begin{enumerate}
\item The invariant is wrong. In this case, the invariant is either too strong or too weak. If it's too strong, it means that you're asking too much to your system. You want your system to verify some invariant but it can't. For example, the following is too strong for the savings account system:
\[
sa \in NIC \rel \num \land pfun(sa) \land \forall x,y . (x \mapsto y \in sa \implies 0 < y)
\]
If it's too weak it means that you're allowing some operations to be called from states they don't expect to be called. For example, the following is too weak for the birthday book:
\[
birthday \in NAME \pfun DATE
\]
\item The operation is wrong. The most common situation is to have a weaker precondition than needed. For example, the following specification of  \textbf{addBirthday} has a precondition making the operation to fail to verify $birthday \in NAME \pfun DATE$:
\[
msg \leftarrow \textbf{addBirthday} (name, date) \defs \\
\t1\bpre~~~ name \in NAME \land date \in DATE \\
\t1\bthen known, birthday, msg := known \cup \{name\}, birthday \cup \{name \mapsto date\}, ok \\
\t1 \bend
\]
Can you tell why? Can you provide a counterexample?
\end{enumerate}

In order to see how \setlog behaves when it fails to prove an invariance lemma, let's assume that the invariant for the birthday book is just: \verb+pfun(Birthday)+. In this case the invariance lemma for \verb+addBirthday+ is as follows:
\begin{Verbatim}
neg(
  pfun(B) &
  addBirthday(K,B,N,C,K_,B_,M) implies
  pfun(B_)
).
\end{Verbatim}
When \setlog is asked to solve the above formula the answer is the following:
\begin{Verbatim}
B = {[N,_N2]/_N1},  
K_ = {N/K},  
B_ = {[N,C],[N,_N2]/_N1},  
M = ok
Constraint: pfun(_N1), comppf({[N,N]},_N1,{}), N nin K, C neq _N2
\end{Verbatim}
As the above formula is satisfiable (which means that the formula inside \verb+neg+ isn't a theorem), \setlog returns a solution that, in this case, is read as a \emph{counterexample}. That is, \setlog returns an assignment of values to variables showing that \verb+addBirthday+ doesn't preserve the invariant. 

By analyzing the counterexample we can discover why \verb+addBirthday+ fails to preserve the invariant giving us the chance to fix the error. The first thing we can do to analyze the counterexample is to replace all the set variables by the empty set\footnote{Except those at the left-hand side of the equalities.}. After a little bit of simplification we obtain:
\begin{Verbatim}
B = {[N,_N2]},
K = {},
K_ = {N},  
B_ = {[N,C],[N,_N2]},  
M = ok
Constraint: C neq _N2
\end{Verbatim}
Observe that \setlog considers executing \verb+addBirthday+ with \verb+B = {[N,_N2]}+ and \verb+K = {}+. This clearly violates $\dom(birthday) = known$. Actually, if we add this condition to the invariance lemma, \setlog returns \verb+no+.
\begin{Verbatim}
neg(
  pfun(B) & dom(B,K) &                     %%% new condition
  addBirthday(K,B,N,C,K_,B_,M) implies
  pfun(B_)
).
\end{Verbatim}
Clearly, now \setlog can't execute \verb+addBirthday+ from a state not verifying $\dom(birthday) = known$.

Recall that in Section \ref{inv} we said that the B invariant can be encoded in \setlog as several clauses (one for each conjunct in the \binv section). In this case, \setlog may fail to prove some invariance lemmas because it needs some of the other invariants as hypothesis. Think that if we separate the invariant of the birthday book in two clauses as we suggest at the end of Section \ref{inv}, \setlog won't be able to prove that \verb+addBirthday+ preserves \verb+pfun(Birthday)+ for the same reason analyzed above. The missing hypothesis can be manually conjoined to the invariance lemmas generated by VCG.

\vfill

\begin{tcolorbox}[%
  toprule=2mm,
  before skip=10pt plus 2pt,
  after skip=10pt plus 2pt,
  title=\textsc{Forgrams}]
What is a \emph{forgram}? Forgram is a portmanteau word resulting from the combination of \emph{for}mula and pro\emph{gram}. A forgram is a piece of code that enjoys the \emph{formula-program duality}. In other words, a forgram is a piece of code that can be used as a formula \emph{and} as a program. In Section \ref{simulacion} we showed that \setlog code can be executed as a program; and in Section \ref{pruebas} we showed that \setlog code can be used as a formula. In \setlog engineers write forgrams, instead of plain programs.
\end{tcolorbox}

\vfill

\begin{tcolorbox}[%
  toprule=2mm,
  before skip=10pt plus 2pt,
  after skip=10pt plus 2pt,
  title=\textsc{Mathematics in Software Development}]
If now most of you are convinced that mathematics is an essential tool for software development, then this course has achieved its objectives.
\end{tcolorbox}

\vfill

\bibliographystyle{plain}
\bibliography{/home/mcristia/escritos/biblio}

\pagebreak

\section{Exercises}

\begin{tcolorbox}[%
  toprule=2mm,
  before skip=10pt plus 2pt,
  after skip=10pt plus 2pt]
Unless stated differently, the proofs indicated in these exercises must be done with \setlog.
\end{tcolorbox}

\begin{enumerate}
  \item\label{ex:sa} Implement in \setlog the following operations of the B specification of the savings account system.
  \begin{enumerate}[ref=\theenumi(\alph*)]
    \item\label{ex:sa:open} Open an account
    \item\label{ex:sa:dep} Deposit money in an account
    \item\label{ex:sa:with} Withdraw money from an account
    \item Query the current balance of an account
    \item Close an account
  \end{enumerate}
  \item\label{ex:trans1} Write it in \setlog the operation specified in exercise 40 of IBM\footnote{Introduction to the B-Mehtod}.
  \item\label{ex:trans2} Concerning exercise \ref{ex:trans1}, can you write a \setlog clause that reuses the clauses defined in exercises \ref{ex:sa:dep} and \ref{ex:sa:with}?
  \item\label{ex:cover1} Run basic simulations that simulate all the disjuncts of all operations implemented in exercise \ref{ex:sa}.
  \item Can you run symbolic simulations on the prototype developed in exercise \ref{ex:sa}? Justify. If you can, do it and analyze the results. For the operations you think you can't, what are your options?
  \item\label{ex:cover2} In the operation of the exercise \ref{ex:sa:open} we have the following abstract assignment:
  \[
  sa := sa \cup \{(n?,0)\}
  \]
  Say we aren't sure this is the right statement. Then, we can simulate the operation with different values to try to decide if the predicate is the right one or not.
  
  To this end we will consider the following \emph{partition} for expressions of the form $S \cup T$.
\[
\begin{array}{ll}
S = \emptyset, T = \emptyset & S \neq \emptyset, T \neq \emptyset, S \subset T \\
S = \emptyset, T \neq \emptyset & S \neq \emptyset, T \neq \emptyset, T \subset S \\
S \neq \emptyset, T = \emptyset & S \neq \emptyset, T \neq \emptyset, T = S \\
S \neq \emptyset, T \neq \emptyset, S \cap T = \emptyset \hspace{1cm} & S \neq \emptyset, T \neq \emptyset, S \cap T \neq \emptyset, S \not\subseteq T, T \not\subseteq S, S \neq T
\end{array}
\]

  How would you do to simulate the \setlog implementation of exercise \ref{ex:sa:open} taking this partition as a reference? Once you have found the method, run the simulations.
  \item\label{ex:systematic} Can you think in a systematic way of generating simulations to do what we asked to do in exercises \ref{ex:cover1} and \ref{ex:cover2}?
  \item Specify in B an operation that opens several accounts at once. Then, translate it into \setlog. Finally, apply what you've learned in exercise \ref{ex:systematic}.
  \item\label{ex:neg} Write in \setlog the following formulas.
  \begin{enumerate}
    \item $\lnot x \in (A \cup B)$
    \item $\lnot (x \in A \land x \in B)$
    \item $\lnot A = B \cap C$
    \item $\lnot (A \cup B = B \cup A)$
    \item $\lnot (A \cap B = \emptyset \implies A = A \setminus B)$
    \item $\lnot (A \subseteq B \implies A \dres R \subseteq B \dres R)$
  \end{enumerate}
  \item Execute in \setlog the formulas of exercise \ref{ex:neg}. Explore all the solutions returned by the tool. Explain why \setlog returns that.
  \item Do exercise 12 of IBM in \setlog.
  \item Prove the results of the following exercises of IBM: 8, 9, 10, 15, 19-24.
  \item Prove that the two clauses defined in exercises \ref{ex:trans1} and \ref{ex:trans2} are equivalent.
\end{enumerate}

\vfill
\pagebreak

\appendix

\section{\label{ap:bb}The \setlog forgram of the birthday book}

\begin{Verbatim}
variables([Known,Birthday]).

def_type(bb,rel(name,date)).
def_type(kn,set(name)).
def_type(msg,enum([ok,nameExists])).

invariant(birthdayBookInv).
dec_p_type(birthdayBookInv(kn,bb)).
birthdayBookInv(Known,Birthday) :- dom(Birthday,Known) & pfun(Birthday).

dec_p_type(n_birthdayBookInv(kn,bb)).
n_birthdayBookInv(Known,Birthday) :- neg(dom(Birthday,Known) & pfun(Birthday)).

initial(birthdayBookInit).
dec_p_type(birthdayBookInit(kn,bb)).
birthdayBookInit(Known,Birthday) :- Known = {} & Birthday = {}.

operation(addBirthday).
dec_p_type(addBirthday(kn,bb,name,date,kn,bb,msg)).
addBirthday(Known,Birthday,Name,Date,Known_,Birthday_,Msg) :-
  (Name nin Known &
   un(Known,{Name},Known_) &
   un(Birthday,{[Name,Date]},Birthday_) &
   Msg = ok
  or
   Name in Known &
   Known_ = Known &
   Birthday_ = Birthday &
   Msg = nameExists
  ).

operation(findBirthday).
dec_p_type(findBirthday(kn,bb,name,date,kn,bb)).
findBirthday(Known,Birthday,Name,Date,Known,Birthday) :-
  Name in Known & applyTo(Birthday,Name,Date).

operation(remind).
dec_p_type(remind(kn,bb,date,kn,kn,bb)).
remind(Known,Birthday,Today,Cards,Known,Birthday) :-
  rres(Birthday,{Today},M) & dom(M,Cards) & dec(M,bb).
\end{Verbatim}

\section{\label{ap:bb-vcg}File generated by VCG for the birthday book}

\begin{verbatim}
% Verification conditions for bb.pl

% Run check_vcs_bb to see if the program verifies all the VCs

:- notype_check.

:- consult('bb.pl').

birthdayBookInit_sat_birthdayBookInv :-
  birthdayBookInit(Known,Birthday) &
  birthdayBookInv(Known,Birthday).

addBirthday_is_sat :-
  addBirthday(Known,Birthday,Name,Date,Known_,Birthday_,Msg) & 
  [Known,Birthday] neq [Known_,Birthday_].

addBirthday_pi_birthdayBookInv :-
  neg(
    % here conjoin other invariants as hypothesis if necessary
    birthdayBookInv(Known,Birthday) &
    addBirthday(Known,Birthday,Name,Date,Known_,Birthday_,Msg) implies
    birthdayBookInv(Known_,Birthday_)
  ).

findBirthday_is_sat :-
  findBirthday(Known,Birthday,Name,Date,Known,Birthday).

findBirthday_pi_birthdayBookInv :-
  % findBirthday doesn't change birthdayBookInv variables
  neg(true).

remind_is_sat :-
  remind(Known,Birthday,Today,Cards,Known,Birthday).

remind_pi_birthdayBookInv :-
  % remind doesn't change birthdayBookInv variables
  neg(true).

check_sat_vc(VCID) :-
  write('\nChecking ') & write(VCID) & write(' ... ') &
  ((call(VCID) & write_ok)!
   or
   write_err
  ).

check_unsat_vc(VCID) :-
  write('\nChecking ') & write(VCID) & write(' ... ') &
  ((call(naf(VCID)) & write_ok)!
   or
   write_err
  ).

write_ok :-
  prolog_call(ansi_format([bold,fg(green)],'OK',[])).

write_err :-
  prolog_call(ansi_format([bold,fg(red)],'ERROR',[])).

check_vcs_bb :-
  check_sat_vc(birthdayBookInit_sat_birthdayBookInv) &
  check_sat_vc(addBirthday_is_sat) &
  check_sat_vc(findBirthday_is_sat) &
  check_sat_vc(remind_is_sat) &
  check_unsat_vc(addBirthday_pi_birthdayBookInv) &
  check_unsat_vc(findBirthday_pi_birthdayBookInv) &
  check_unsat_vc(remind_pi_birthdayBookInv) &
  true.

:- nl & 
   prolog_call(ansi_format([bold,fg(green)],
               'Type checking has been deactivated.',[])) & 
   nl & nl.

:- nl &
   prolog_call(ansi_format([bold,fg(green)],
               'Call check_vcs_bb_b_inv to run the verification conditions.',
               [])) &
   nl & nl.
\end{verbatim}

\end{document}